\documentclass[]{aastex631}
\usepackage{subfigure}

\usepackage[normalem]{ulem}
\usepackage{CJKutf8}
\usepackage[inkscapelatex=false]{svg}

\def\amin{\ifmmode^{\prime}\else$^{\prime}$\fi}
\def\asec{\ifmmode^{\prime\prime}\else$^{\prime\prime}$\fi}

\def\simgt{\lower.5ex\hbox{$\; \buildrel > \over \sim \;$}}
\def\simlt{\lower.5ex\hbox{$\; \buildrel < \over \sim \;$}}

\newcommand{\fluxcgs}{erg~s$^{-1}$~cm$^{-2}$}
\newcommand{\lumcgs}{erg~s$^{-1}$}

\newcommand\chandra{{\it Chandra}}

\newcommand\xmm{{\it XMM-Newton}}
\newcommand\XMM{{\it XMM-Newton}}

\newcommand\nustar{{\it NuSTAR}}
\newcommand\nicer{{\it NICER}}

\newcommand\src{ZTF J1851} 

\begin{document}

\title{Broadband X-ray observations of the periodic optical source ZTF J185139.81+171430.3 and its identification as a massive intermediate polar}

\author[0009-0003-4151-7735]{Ren Deng}
\affiliation{Columbia Astrophysics Laboratory, Columbia University, New York, NY 10027, USA}

\author[0000-0002-9709-5389]{Kaya Mori}
\affiliation{Columbia Astrophysics Laboratory, Columbia University, New York, NY 10027, USA}

\author[0009-0009-6455-3804]{Eric Miao}
\affiliation{Columbia Astrophysics Laboratory, Columbia University, New York, NY 10027, USA}

\author[0000-0002-6653-4975]{Gabriel Bridges}
\affiliation{Columbia Astrophysics Laboratory, Columbia University, New York, NY 10027, USA}

\author[0000-0002-3681-145X]{Charles J. Hailey}
\affiliation{Columbia Astrophysics Laboratory, Columbia University, New York, NY 10027, USA}

\author[0000-0002-7004-9956]{David A. H. Buckley}
\affiliation{South African Astronomical Observatory, P.O Box 9, Observatory, 7935 Cape Town, South Africa}
\affiliation{Department of Astronomy, University of Cape Town, Private Bag X3, Rondebosch 7701, South Africa}
\affiliation{Department of Physics, University of the Free State, PO Box 339, Bloemfontein 9300, South Africa}

\author[0000-0001-8722-9710]{Gavin Ramsay}
\affiliation{Armagh Observatory and Planetarium, College Hill, Armagh, BT61 9DG, UK}

\author[0009-0004-3067-2227]{Dan Jarvis}
\affiliation{Astrophysics Research Cluster, School of Mathematical and Physical Sciences, University of Sheffield, Sheffield S3 7RH, UK}

\begin{abstract}

We present X-ray observations of the periodic optical source ZTF J185139.81+171430.3 (hereafter ZTF J1851) by the \xmm, \nicer\ and \nustar\ telescopes. The source was initially speculated to be a white dwarf (WD) pulsar system due to its short period ($P\sim12$ min) and highly-modulated optical lightcurves. Our observations revealed a variable X-ray counterpart extending up to 40 keV with an X-ray luminosity of $L_X \sim 3\times10^{33}$~\lumcgs\ (0.3--40 keV). 
Utilizing timing data from \xmm\ and \nicer, we detected a periodic signal at $P_{\rm spin}=12.2640(7)\pm0.0583$ min with $>6\sigma$ significance. The pulsed profile displays $\sim 25\%$ and $\sim10$\% modulation in the 0.3--2 and 2--10 keV bands, respectively. 
Broadband X-ray spectra are best characterized by an absorbed optically-thin thermal plasma model with $kT \approx 25$ keV and a Fe K-$\alpha$ fluorescent line at 6.4 keV. 
The bright and hard X-ray emission rules out the possibility of a WD pulsar or ultra-compact X-ray binary. 
The high plasma temperature and Fe emission lines suggest that ZTF J1851 is an intermediate polar spinning at 12.264 min. 
We employed an X-ray spectral model composed of the accretion column emission and X-ray reflection to fit the broadband X-ray spectra. Assuming spin equilibrium between the WD and the inner accretion disk, we derived a WD mass range of $M_{\rm WD}=(1.07\rm{-}1.32)M_{\odot}$ exceeding the mean WD mass of IPs ($\langle M_{\rm WD} \rangle = 0.8 M_\odot)$. Our findings illustrate that follow-up broadband X-ray observations could provide unique diagnostics to elucidate the nature of periodic optical sources anticipated to be detected in the upcoming Rubin all-sky optical surveys.  

\end{abstract}

\section{Introduction} \label{sec:intro}

The Zwicky Transient Facility (ZTF) is one of the leading optical survey programs in contemporary time-domain astrophysics \citep{bellm_zwicky_2019}. Since 2018, ZTF has provided wide-field coverage of the entire northern sky and discovered a number of supernovae, novae and other transient objects. The vast ZTF lightcurve data in the g, r and i bands also serve as a search engine for pulsating stars and binaries by detecting periodic signals. The high-cadence optical monitoring of more than a billion Galactic sources revealed a rare class of short-period objects with P $\simlt$ 1 hr. Some of these short-period objects turned out to be ultra-compact binaries (UCBs) with a white dwarf (WD) or neutron star (NS) primary, in addition to fast-spinning WDs. For example, the discovery of a 62-min-orbit black widow pulsar, where a millisecond pulsar ablates its companion via pulsar wind, highlights the potential of finding exotic binaries with ongoing optical surveys including the upcoming Rubin observatory Legacy Survey of Space and Time (LSST) \citep{burdge_62-minute_2022}.

\begin{figure}[ht]
    \noindent
    \includegraphics[width=\textwidth]{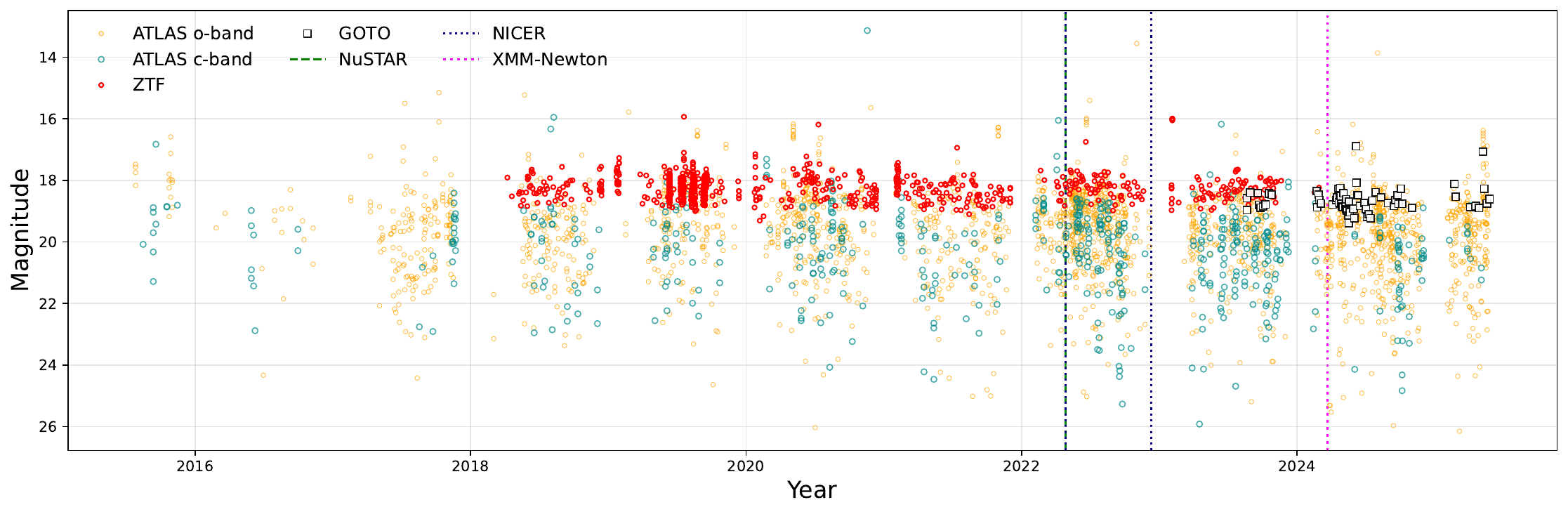}

    \caption{Long-term optical lightcurve of ZTF J1851. Occasional day-long flares have been observed. The X-ray observation dates are marked by vertical lines.}
    \label{fig:lightcurve} 
\end{figure}

Among the variable optical sources detected by ZTF, ZTF J1851 displayed a $P = 12.37$ min periodicity with high modulation of $\sim 0.8$ mag \citep{kato_ztf_2021}. In addition, several day-long outbursts with an increase in luminosity of $\sim$ 2 mag have been detected (see Figure \ref{fig:lightcurve}) by ZTF, GOTO, and ATLAS.
In Figure \ref{fig:lightcurve}, we also show L band (400-700 nm) photometry obtained using the GOTO all-sky survey which consists of 32 0.4 m telescopes located in La Palma in the Canaries and Siding Spring in Australia (\cite{2024SPIE13094E..1XD}, \cite{steeghs_gravitational-wave_2022}). 
In addition, optical observations by ATLAS \citep{tonry_atlas_2018} are displayed in Figure \ref{fig:lightcurve}.

It is unknown whether the 12-minute periodicity represents a spin or orbital period, each of which would correspond to a different source type for ZTF J1851. In the spin period case, ZTF J1851 may be a cataclysmic variable (CV), an interacting binary harboring a WD accreting material from a Roche-lobe filling late-type main sequence companion. Magnetic CVs, either of the polar or intermediate polar (IP) types, emit copious X-rays with $L_{\rm X} \gtrsim 10^{33}$~erg\,s$^{-1}$ from their accretion columns in the form of thermal bremsstrahlung emission and atomic lines \citep{mukai_x-ray_2017}. Alternatively, the highly modulated, coherent periodicity in the optical band suggests that ZTF J1851 could be another WD pulsar spinning at $P=12.37$ min. Unlike magnetic CVs, the only two known WD pulsar systems, AR Sco and J191213.72-441045.1, exhibit faint X-ray emission ($L_X \sim 10^{30}$ erg\,s$^{-1}$) most likely of thermal origin \citep{Schwope2023}. For AR Sco, pulsed non-thermal X-ray emission below 2 keV was reported \citep{takata_x-ray_2021}, consistent with magnetic dipole emission, rather than accretion. 

In the orbital period case, ZTF J1851 could be a UCB containing either a WD {(AM CVn)} or an NS {(UCXB)}. AM CVn stars are close binary systems with $P_{orb} \lesssim$ 1 hr, composed of a WD and a low-mass companion. {Apart from the direct impact systems, HM Cnc and V407 Vul, which show
strong soft X-rays with $kT_{\rm{BB}}<0.1$ keV, AM CVn binaries typically show
thermal bremmstrahlung emission with $kT_{\rm{max}} \sim5-6$ keV} \citep{ramsay_xmm-newton_2005}. On the other hand, NS-UCBs are accreting NS binaries with tight orbits ($P_{orb} \lesssim$ 1 hr). Among $\sim$ 40 known UCXBs and candidates, including 15 sources confirmed through observed type I X-ray bursts, only three sources have $P_{orb} \lesssim$ 30 min \citep{koliopanos_chemical_2021}. 
ZTF J1851 resembles a 12.8-minute orbit UCXB candidate (OGLE-UCXB-01) discovered by the OGLE optical survey and also detected in the X-ray band by \chandra\  \citep{pietrukowicz_discovery_2019}{, with similar X-ray luminosities, high optical modulation, and occasional optical flares}. 
These {known} UCXBs show power-law X-ray spectra with either no or weak Fe K emission lines ($\textup{EW} < 100$ eV) reflected off the accretion disk \citep{koliopanos_chemical_2021}.

Follow-up X-ray observations of ZTF J1851 are important for elucidating the source type. We performed X-ray observations and detected an X-ray counterpart with \nustar, \nicer\ and \xmm\ in 2022--2024. 
In this paper, we present the first X-ray spectral and timing analysis of ZTF J1851 using broadband X-ray data. In \S\ref{sec:x-ray}, we describe the X-ray observations and data processing. In \S\ref{sec:timing}, we describe our X-ray timing analysis, searching for an X-ray periodicity and characterizing X-ray lightcurves. In \S\ref{sec:source}, we describe various source type candidates for ZTF J1851 and identify it as an IP based on its X-ray timing and spectral properties. With phase-averaged and phase-resolved spectral analysis, we constrain several critical physical parameters, including metallicity, plasma temperature and Fe emission lines. {In \S\ref{sec:discussion}, we discuss the implications of our timing and spectral analysis results. }In \S\ref{sec:mass}, we employ our physically-motivated spectral model for constraining the WD mass of ZTF J1851.  Finally, we conclude the paper with future prospects in \S\ref{sec:conclusion}. 

\section{X-ray Observations and Data Reduction} \label{sec:x-ray}

ZTF J1851 was observed by \textit{NuSTAR} (42.6 ks) and \textit{NICER} (9.5 ks) in April 2022, and was subsequently observed by \textit{XMM-Newton} in March 2024 (79 ks). Figure \ref{fig:lightcurve} shows the optical lightcurve of the source, overlaid with vertical lines indicating X-ray observation dates. 
For \textit{NuSTAR}, we processed the data using \texttt{nupipeline} (\cite{harrison_nuclear_2013}). For timing and spectral analysis, we extracted source events from a $r = 12$\asec\ circle around the source position. Background events were extracted from a source-free $r = 107$\asec\ circle region on the same detector chip where the source is located. The \textit{NuSTAR} data was contaminated by background above $35$ keV. We collected $\sim1000$ source counts after background subtraction. For \textit{XMM-Newton} data, we reduced the datasets using SAS 21.0. We utilize \textit{emchain} and \textit{epchain} for GTI filtering for MOS and PN cameras, respectively. After filtering, this leaves an exposure time of about $54$ ks for PN camera and $74$ ks for MOS cameras. We extracted a source region with a circle of radius $r=400$\asec, with an annular region of $r=500-1500$\asec\ used for background extraction. We collected $\sim 7,100$ and $\sim 10,000$ source counts in the combined MOS and PN cameras, respectively. We processed the \nicer\ data using the \texttt{nicerl2} and \texttt{nicerl3-spect} tasks in the \nicer\ Data Analysis Software (\texttt{NICERDAS} version 13). For \nicer\, the \texttt{SCORPEON} background model was used when fitting spectral data. {We discarded the December \nicer\ observations because their photon counts are close to zero. For April observations, we used only observations 5594010103 and 5594010104.}

\begin{deluxetable}{lcccc}[ht]
\tablecaption{X-ray Observations of ZTF J1851}
\label{tab:x-ray}
\tablecolumns{4}
\tablehead{
\colhead{Observation Date} &
\colhead{Telescope} &
\colhead{ObsID} &
\colhead{Exposure (ks)} }
\startdata
2022-04-28 & \textit{NuSTAR} & 30801008002 & 42.6  \\
2022-04-27/28 & \textit{NICER} & $559401010^a$ & 9.5\\
2022-12-11/13 & \textit{NICER} & $559401020^a$ & 5.7\\
2024-03-22 & \textit{XMM-Newton} & 0921690101 & 79 
\enddata
\vspace{-5pt} 
\tablenotetext{a}{\textit{NICER} observations are collected in several successive observations sharing the same obsID prefix.}
\end{deluxetable}

\section{Timing Analysis} \label{sec:timing}

We performed timing analysis on the X-ray observation data using the \texttt{Stingray} (\cite{huppenkothen_stingray_2019}, \cite{bachetti_stingray_2024}) and \texttt{Hendrics} \citep{bachetti_hendrics_2018} software packages. First, we applied barycentric corrections to all extracted source events using the SAS command \textbf{barycen} for the \textit{XMM-Newton} data and the HEASOFT command \textbf{barycorr} for the \nustar\ and \nicer\ data. 
\subsection{Periodicity Search} \label{subsec:tables}

\begin{figure}[htbp]
    \centering
    \subfigure{
    \includegraphics[width=0.47\textwidth]{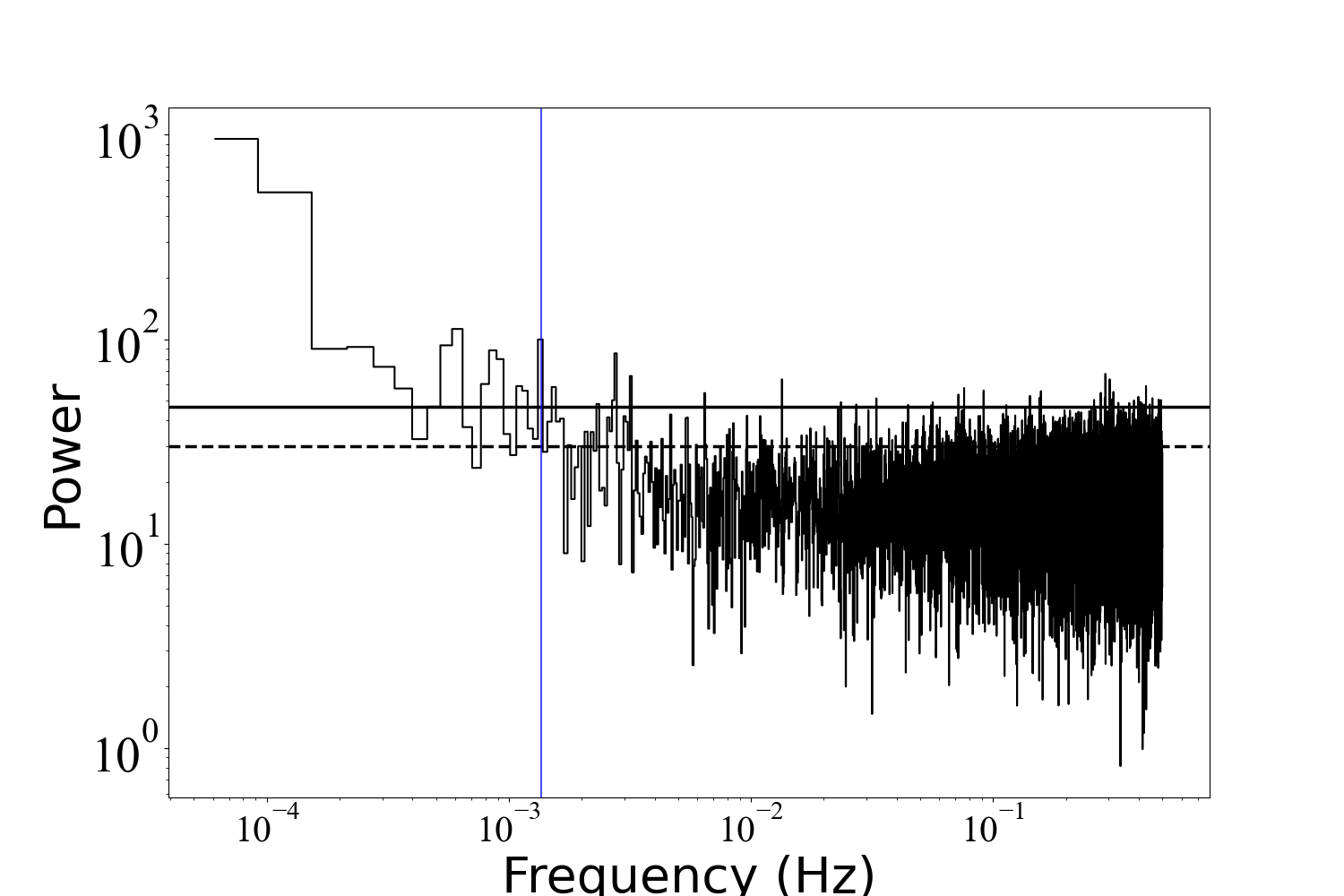}
        \label{fig:first}
    }
    \subfigure{
        \includegraphics[width=0.47\textwidth]{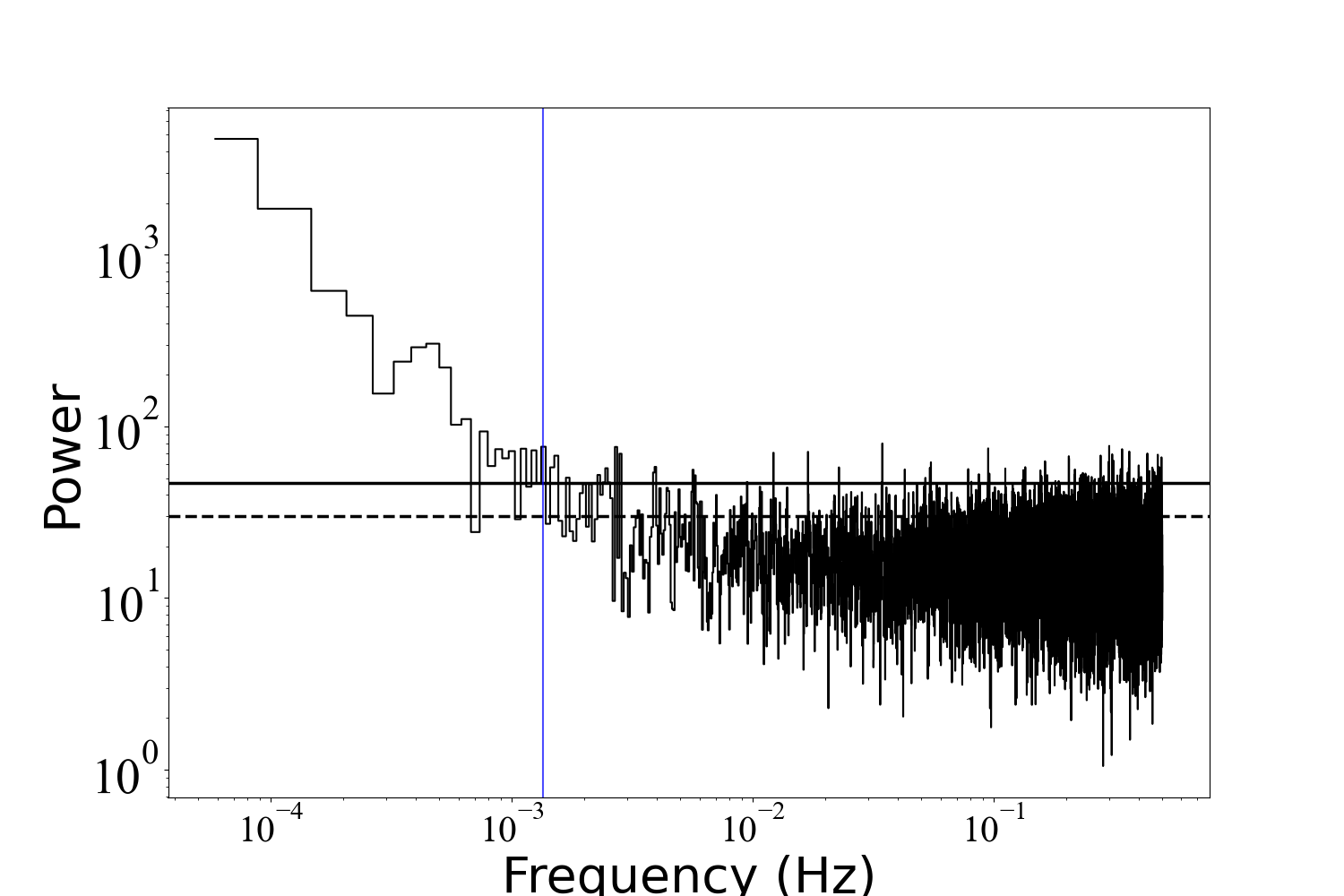}
        \label{fig:second}
    }
    \caption{PDS periodograms obtained from the combined MOS (left) and PN (right) cameras. Blue vertical line marks the known optical periodicity $P=12.37$ mins. Note that both axes are in log scale. The dashed and solid lines denote the $3\sigma$ and $5\sigma$ significance, respectively.}
    \label{fig:PDS}
\end{figure}

We conducted the $Z$-test for $n=1,2,3$ harmonic components using the weighted $Z^2_n$ function through \textbf{z}$\_$\textbf{n}$\_$\textbf{search} in \texttt{Stingray}. We adopted $n = 2$ for the number of harmonics in the $Z^2$ test {for \xmm}. Including higher harmonic components did not improve the $Z^2$ statistics. The power density spectra (PDS) over a broad frequency band are dominated by red noise below $f\simlt 10^{-3}$ Hz, yielding no significant peak detection  (Figure \ref{fig:PDS}). We conducted a refined period search in a narrower frequency range ($f=0.5-2.5$ mHz) around the optical period at $P= 12.37$ min corresponding to $f\approx 1.36$ mHz. {No significant periodicity was detected in the \nustar\ data (3--40 keV), likely due to the lower source counts.} We detected a significant periodic signal ($>6\sigma$) at $P=12.26$ min period below 2 keV in the PN data. The signal was detected with $>8\sigma$ significance in the combined MOS data. 
Above $E = 2$ keV, the periodic signal was detected in the \xmm\ data with lower significance ($\sim5\sigma$). In the narrow-band $Z^2$ data, we fit the 12.26-min peak with a Gaussian and constrain its width to $\Delta f<0.014$ mHz. As the frequency resolution is defined by the reciprocal of the observation time, the measured period $P_{\rm spin}=12.2640(7)\pm0.0583$ min is well-constrained due to the long baseline of 70 ks \xmm\ observation time. 

Notably, this measured X-ray period ($P = 12.26$ min) very significantly differs from the previously known optical period at $P = 12.36$ min. In addition, both the \xmm\ and \nicer\ data showed no noticeable modulation when folded at the optical period of $P = 12.37$ min. After investigating possible aliases, we identified the optical period as simply the one-day alias of the X-ray period. This is expected for single-site (ZTF) data. Thus, we conclude that the X-ray period of $P = 12.26$ min represents the intrinsic periodicity of ZTF J1851. 

For the \textit{NICER} data, we used the \texttt{Hendrics} command \textbf{HENzsearch} to attempt to conduct $Z^2$ tests, with $n = 1$ for the number of harmonics. Inspecting the entire observational lightcurve revealed short GTIs of $\sim 15$ min in length, spaced $
\sim 93 $ min apart. As a result, we also utilized \texttt{astropy's} \textbf{LombScargle} to conduct a floating-point Lomb Scargle test (\cite{scargle_studies_1982}, \cite{lomb_least-squares_1976}) at $20\,s$ per bin. Both pulsation searches gave similar results, with statistically insignificant ($<1\sigma$) peaks detected at the measured X-ray frequency ($\textit{f} \approx 1.36$ mHz). A possible explanation is the dominance of the ISS period and its harmonics ($f \sim n*0.18$ mHz) on both spectra, obscuring the significance of the signal frequency. 

\begin{figure}[htbp]
    \centering
    \subfigure{
    \includegraphics[width=0.47\textwidth]{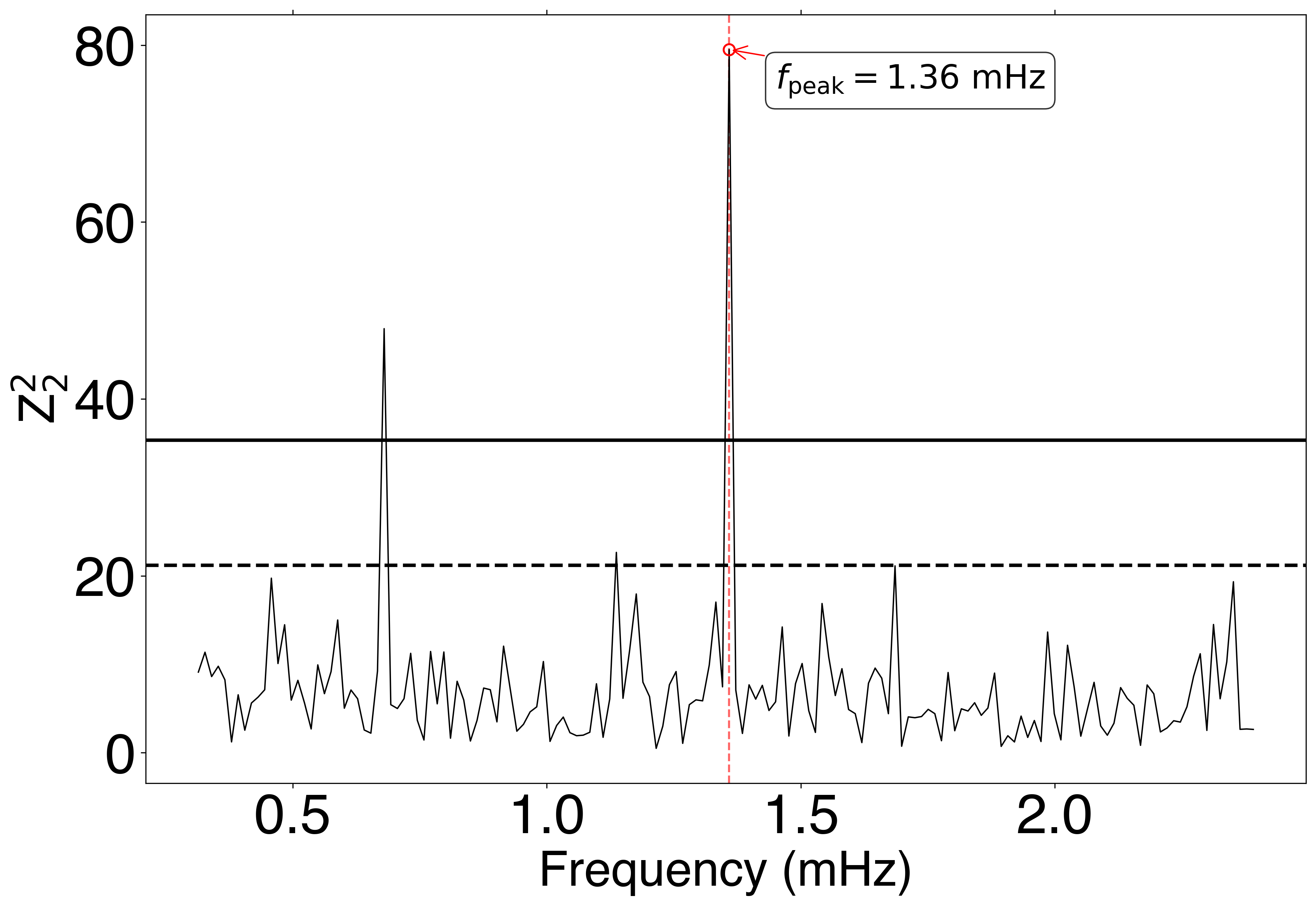}
        \label{fig:first}
    }
    \subfigure{
        \includegraphics[width=0.47\textwidth]{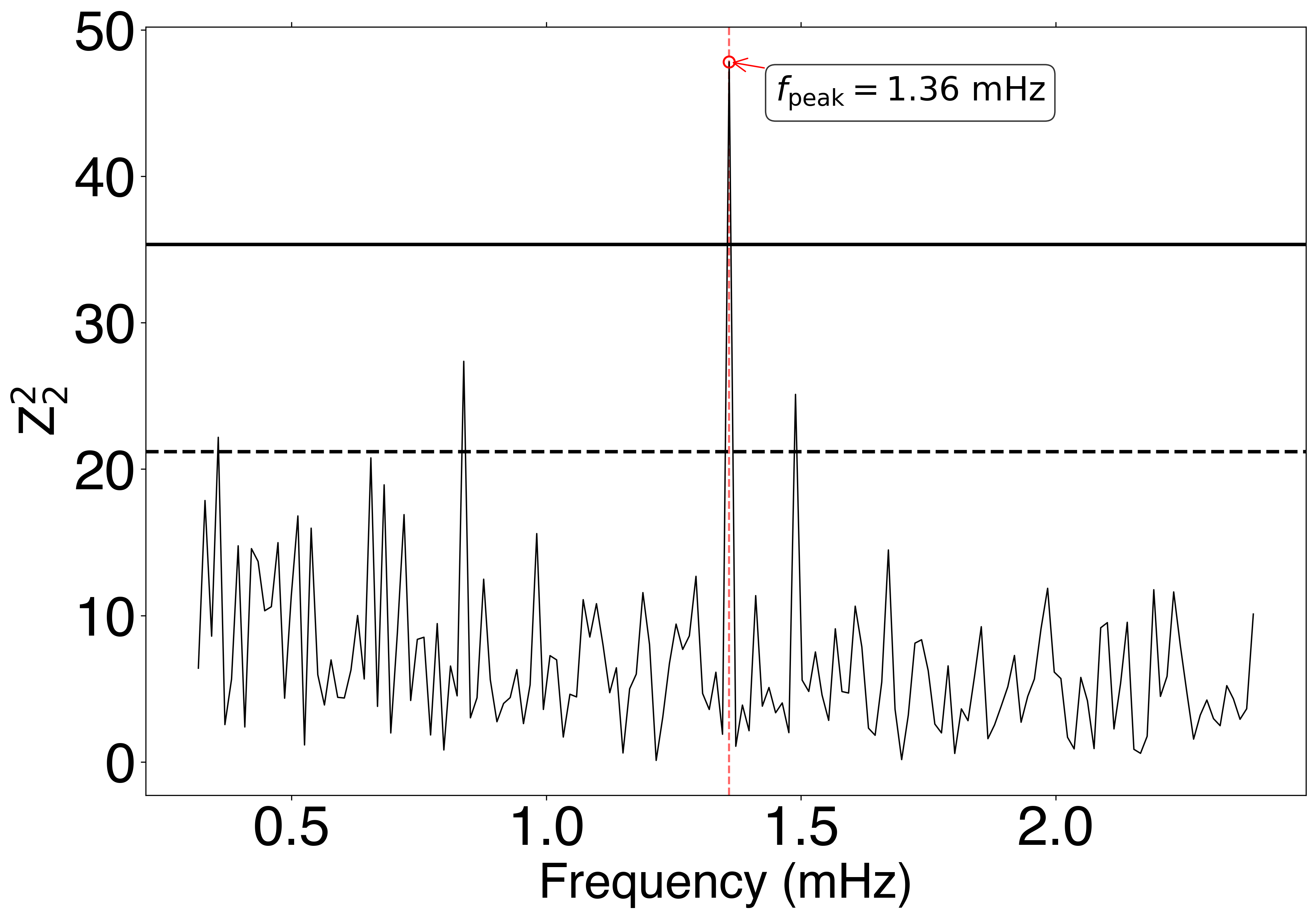}
        \label{fig:second}
    }
    \caption{$Z^2_2$ soft X-ray (0.3-2 keV) periodograms. Left: combined MOS cameras. Right: PN camera. The red vertical line marks the most significant peak. The dashed and solid lines denote the $3\sigma$ and $5\sigma$ significance, respectively.}
    \label{fig:periodograms}
\end{figure}

\subsection{Folded X-ray Lightcurves} 
Using \texttt{Stingray}'s \textbf{fold}$\_$\textbf{events}, we produced exposure-corrected pulsed profiles folded at $P = 12.26$ min for \nicer, PN and MOS data separately. {For \xmm, we subtracted the source  profile with a background profile normalized using the BACKSCAL and AREASCAL values from the spectral files. For \nicer\  lightcurves, we folded the estimated background lightcurve produced by the SCORPEON background model in nicerl3-lc, and subtracted it from the total profile. Subtracting out backgrounds ensure accurate determination of pulsed fractions and hardness ratios. }

\begin{figure}[htbp]
    \centering
    \subfigure{
    \includegraphics[width=0.31\textwidth]{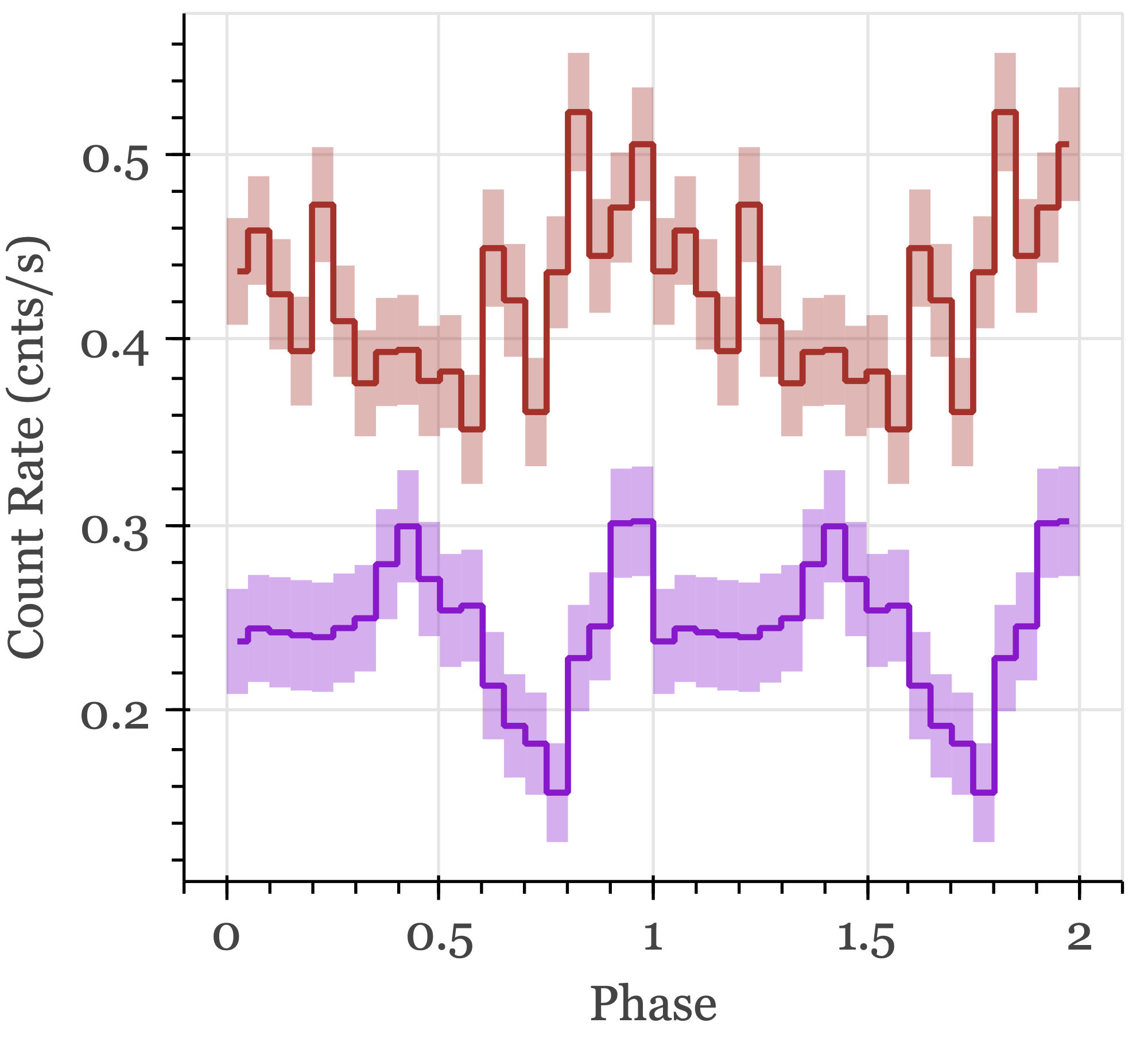}
        \label{fig:second}
    }    
    \subfigure{
    \includegraphics[width=0.31\textwidth]{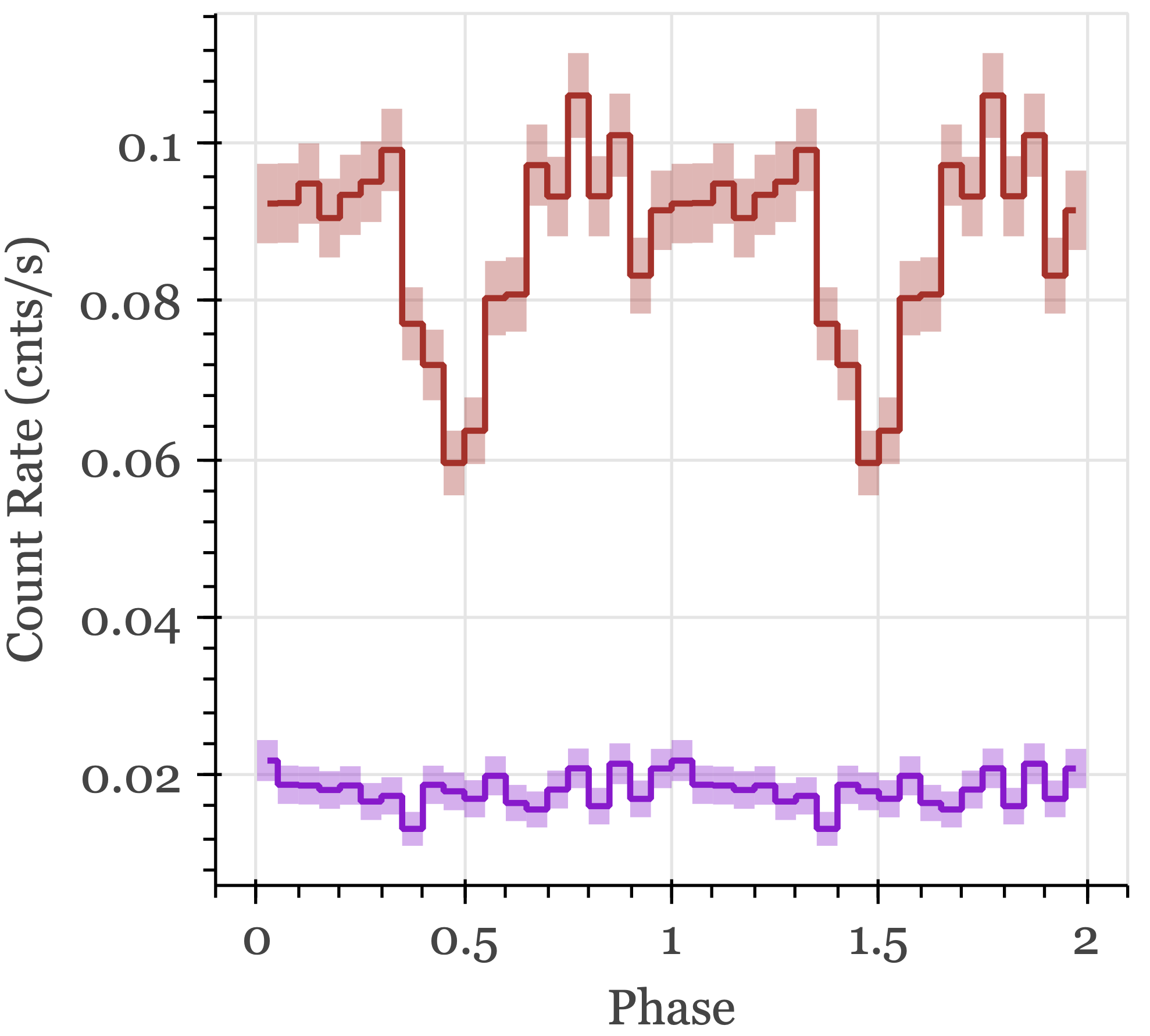}
        \label{fig:second}
    }
    \subfigure{
        \includegraphics[width=0.31\textwidth]{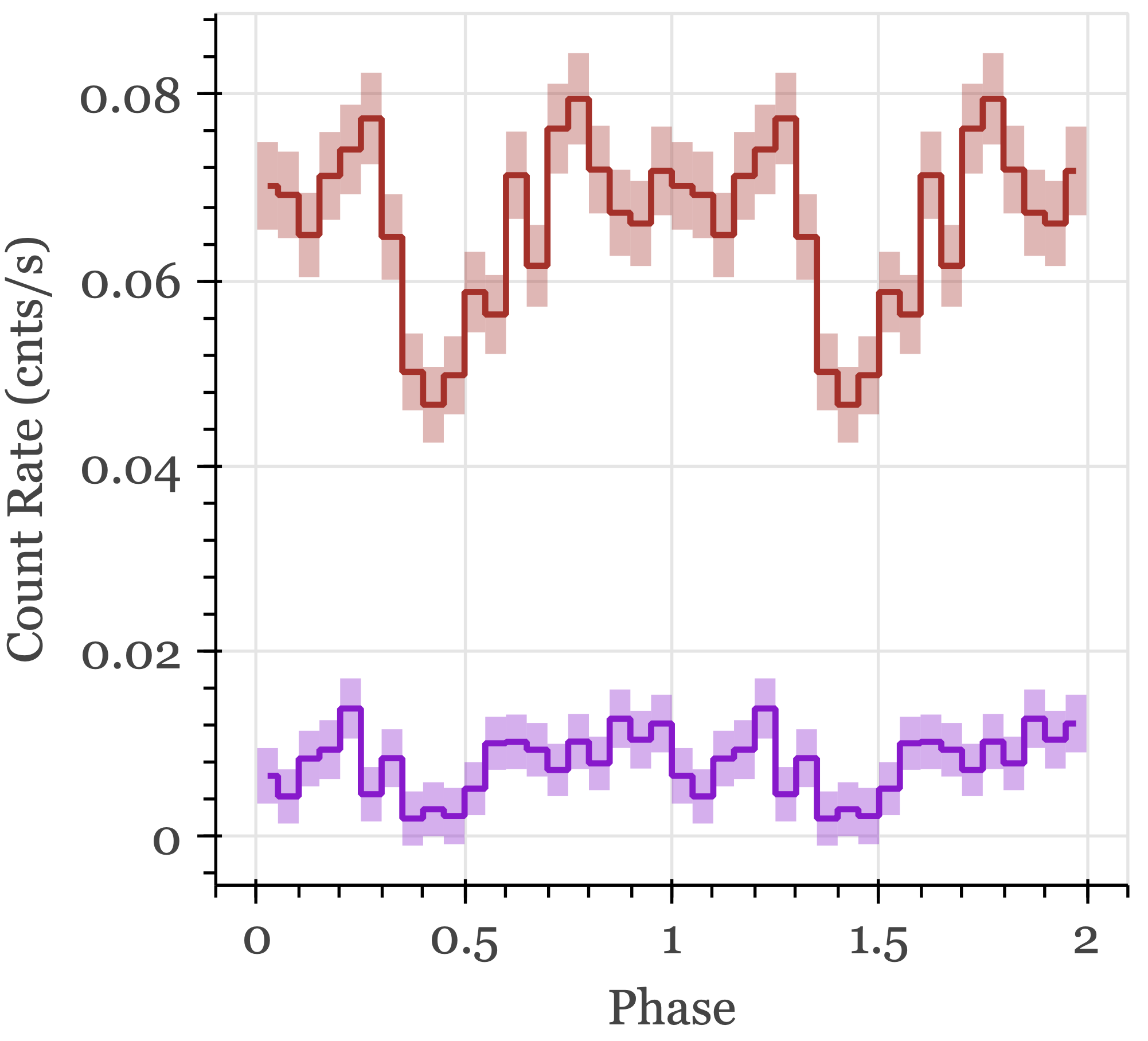}
        \label{fig:third}
    }
    \subfigure{
        \includegraphics[width=0.31\textwidth]{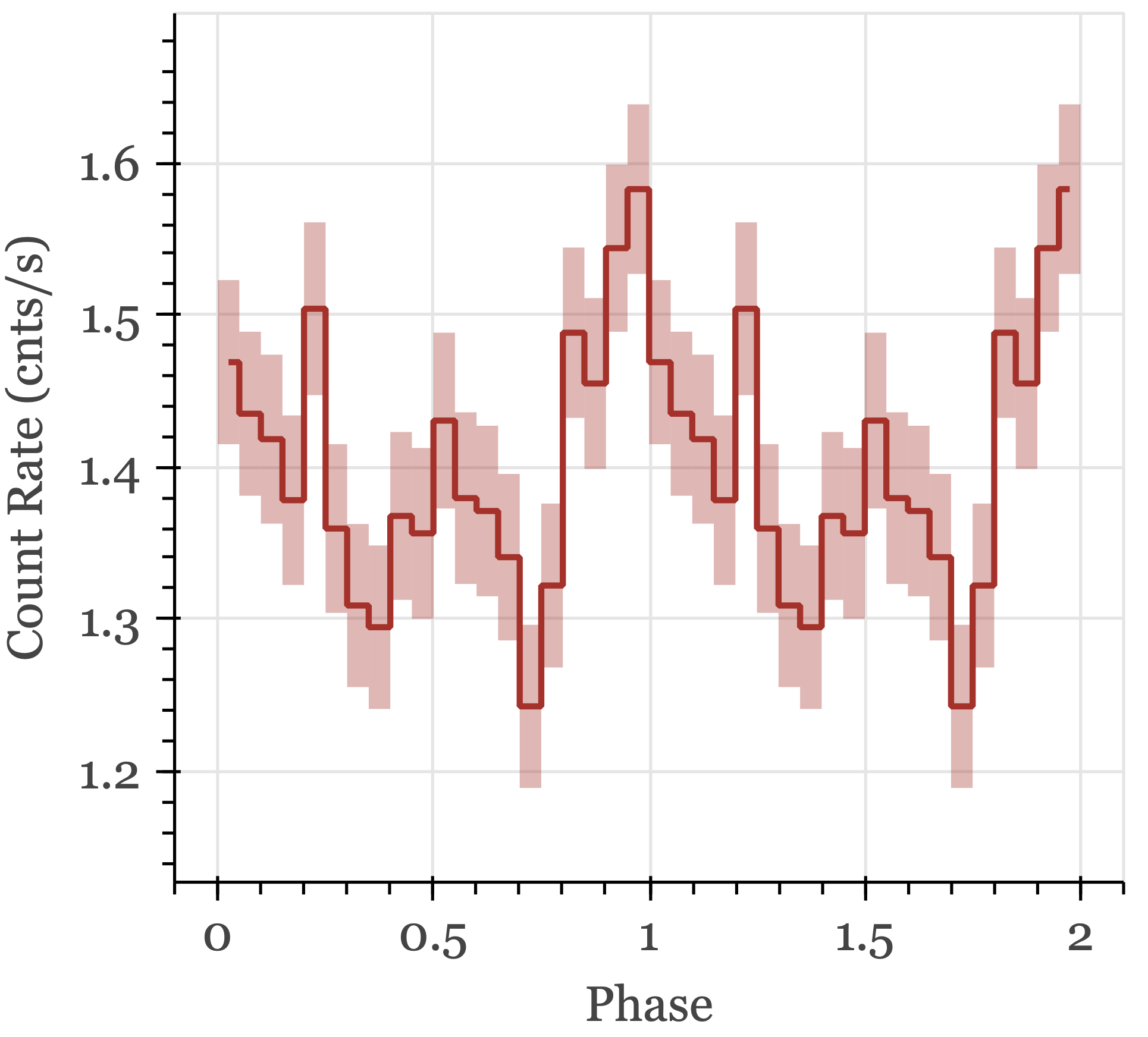}
        \label{fig:first}
    }    
    \subfigure{
    \includegraphics[width=0.31\textwidth]{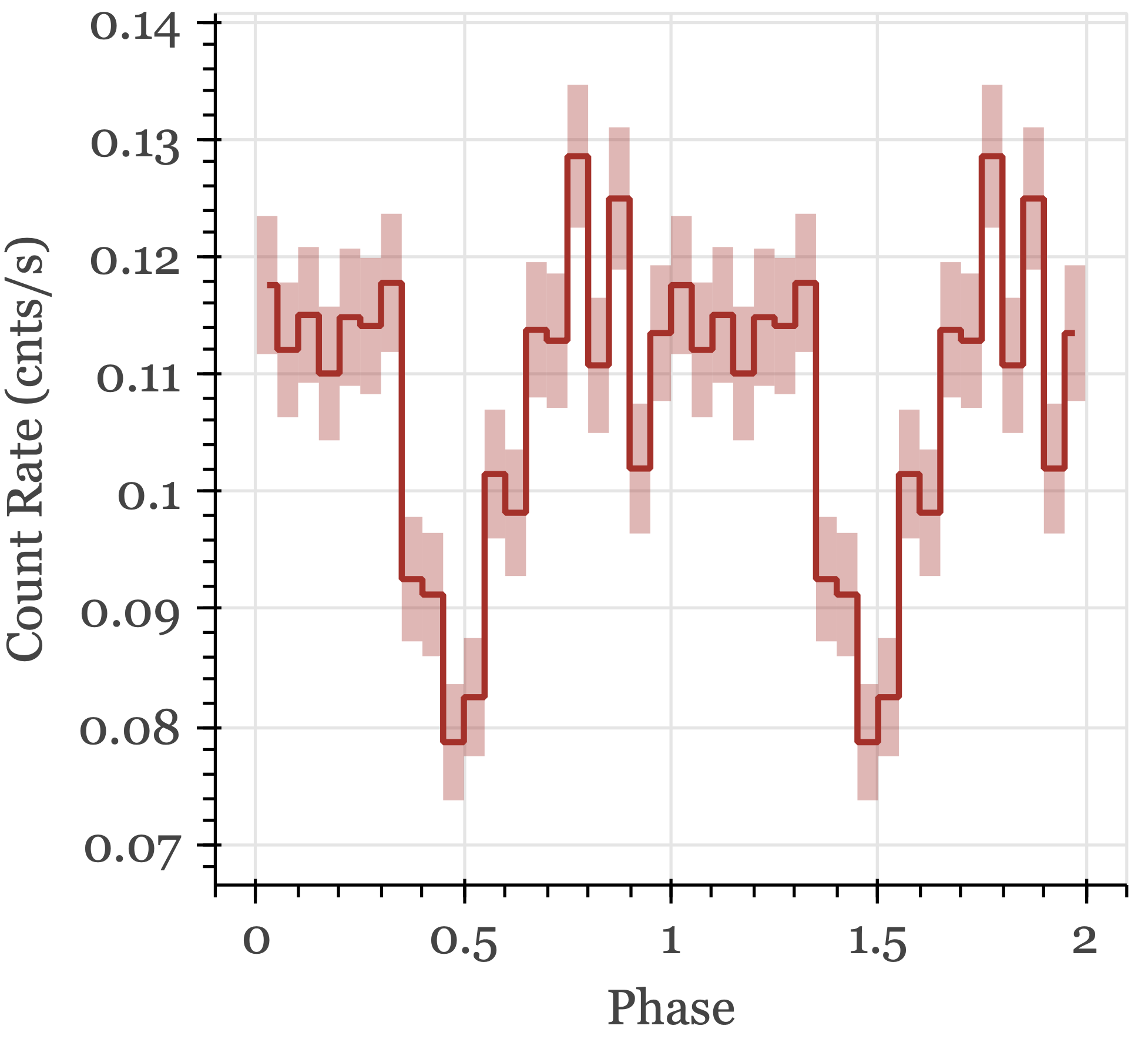}
        \label{fig:second}
    }
    \subfigure{
        \includegraphics[width=0.31\textwidth]{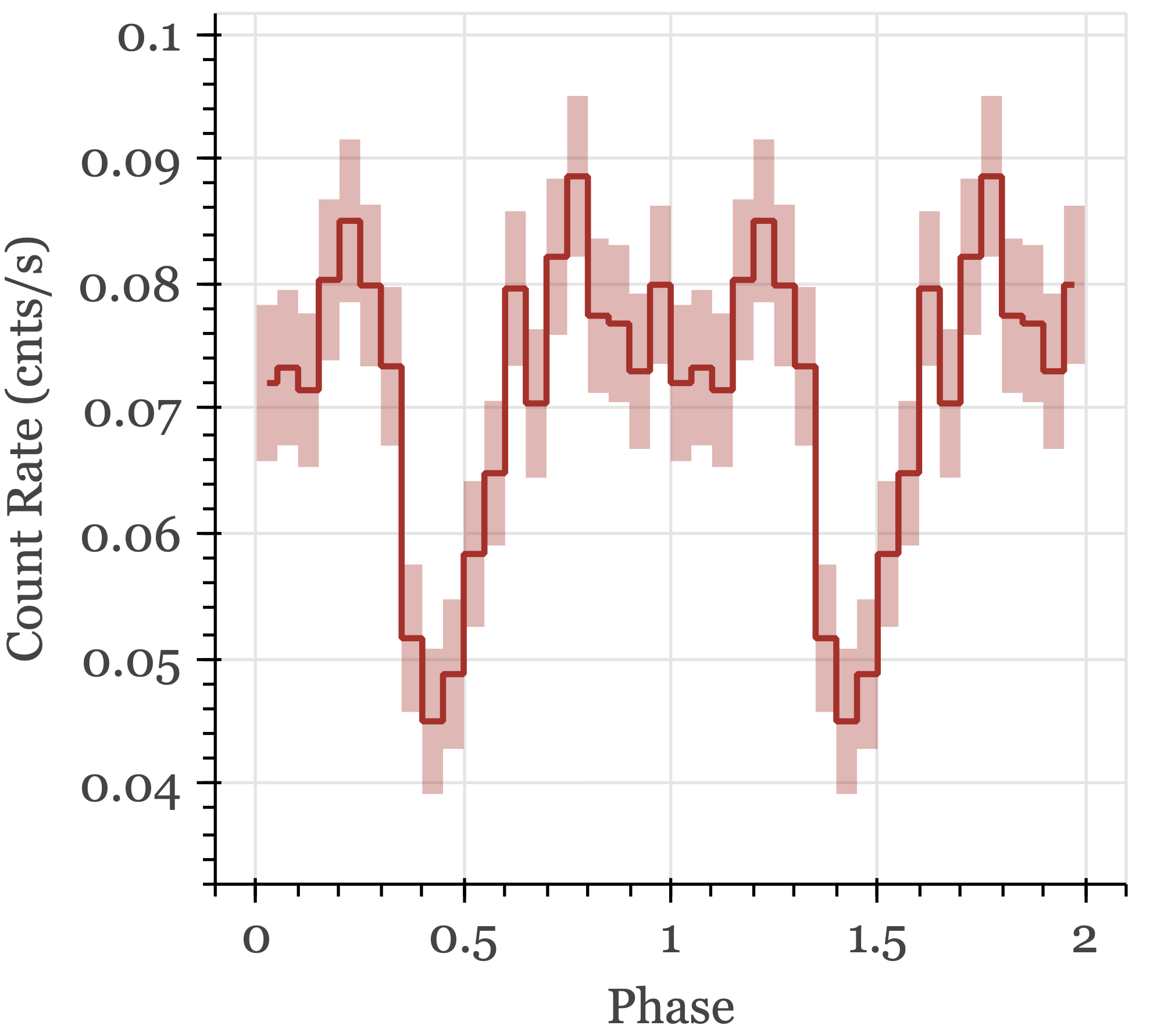}
        \label{fig:third}
    }
    \caption{Folded lightcurve plots in the soft (0.3--4 keV for \xmm\ and 1--4 keV for \nicer\ ; red) and hard (4--10 keV; purple) X-ray band, with 1-$\sigma$ error bars shown (left: \nicer\; middle: XMM-MOS; right: XMM-PN). The bottom three figures show the combined folded lightcurves including photon energy up to $10$ keV. All light curves are folded over the $P=12.26$ min period, phase-aligned, and contain 20 bins per phase. }
    \label{fig:profiles}
\end{figure}

Folded lightcurves in the {$0.3-4$} keV and {$4-10$} keV energy bands are shown in Figure \ref{fig:profiles}, each with 20 bins per phase. {For \nicer\, we used $1-4$ keV for the soft energy band, since photon counts below $1$ keV are dominated by background.} {The pulse fraction is defined as $\frac{a-b}{a+b}$, where $a$ and $b$ are respectively the minimum and maximum of the profile. Soft and hard X-ray lightcurves from \xmm\ exhibit a pulse fraction between 20--25\%, and \nicer\ lightcurves exhibit a comparable pulse fraction of $\sim20$\% in the soft band but a stronger $30$\% in the hard band. Significant periodic modulation in the \nicer\ folded lightcurves demonstrates the existence of a 12.26 min period despite insignificant $Z^2$ and Lomb-Scargle tests.}

\begin{figure}[htbp]
    \centering
    \subfigure{
    \includegraphics[width=0.47\textwidth]{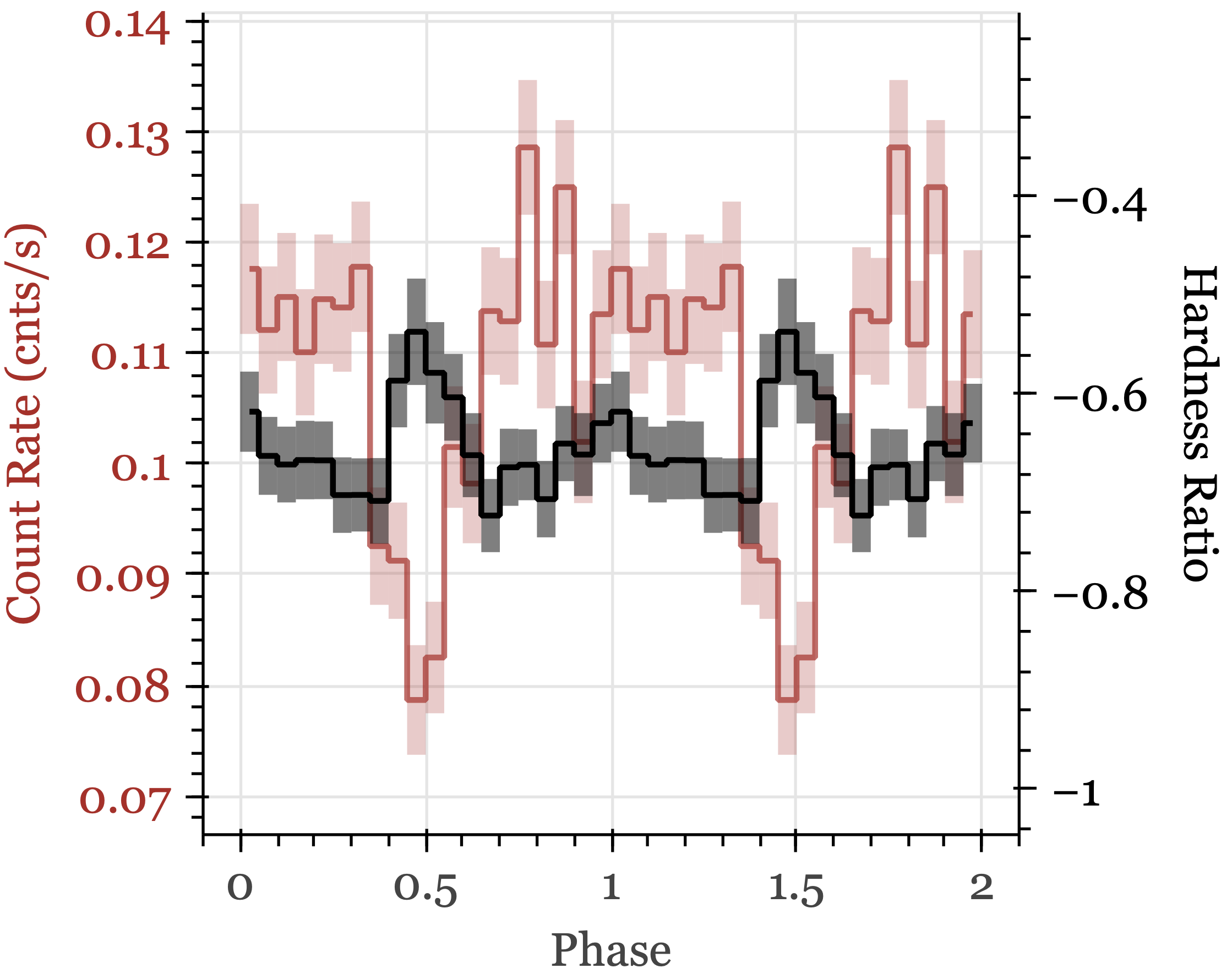}
        \label{fig:first}
    }
    \subfigure{
        \includegraphics[width=0.47\textwidth]{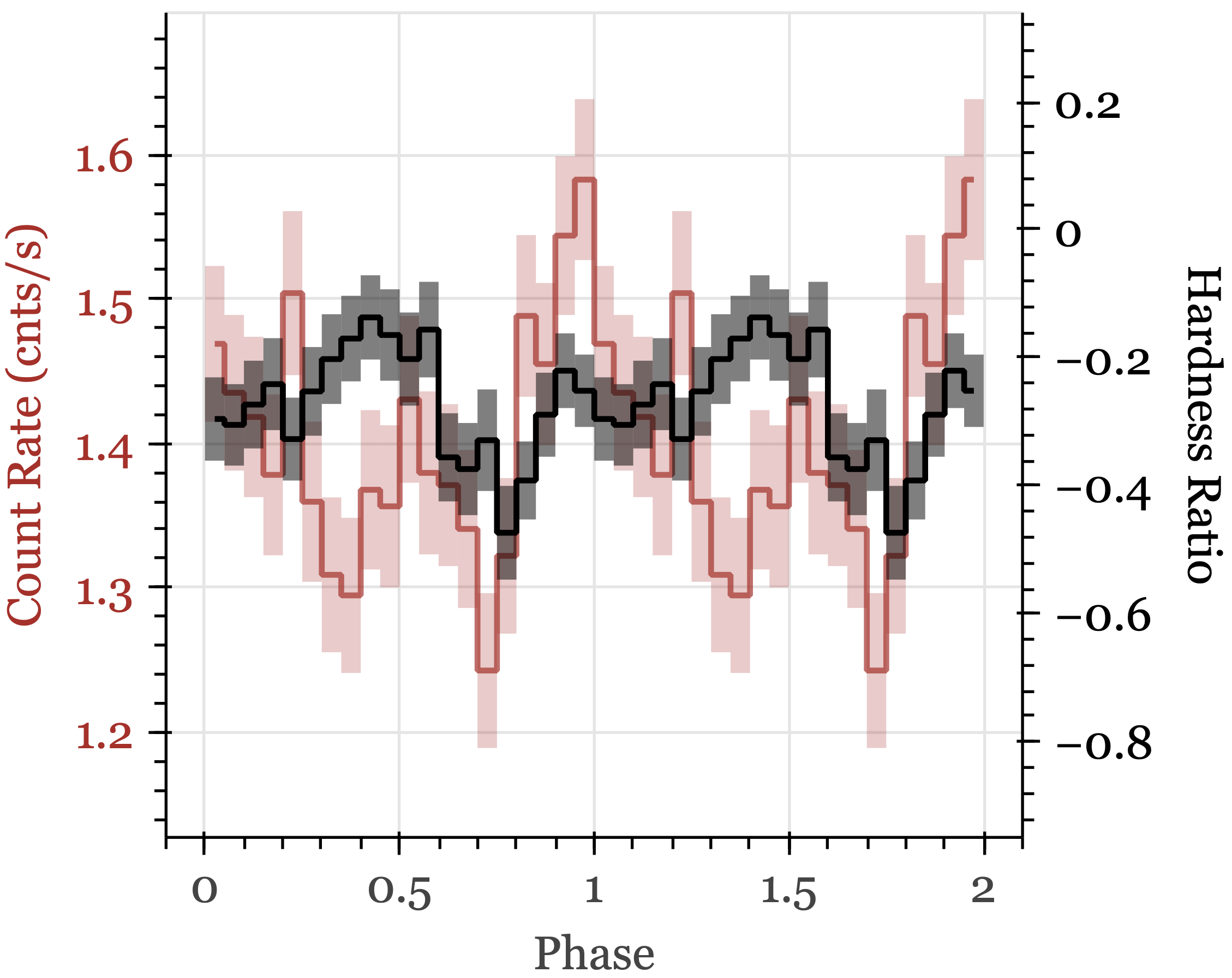}
        \label{fig:third}
    }    
    \caption{{The hardness ratios (black) overlaid with the $0.3-10$ keV (left: XMM-MOS) and $1-10$ keV (right: \nicer) folded lightcurves, showing an anti-correlation. Both are folded at the $P=12.26$ min period. Hardness ratio is defined as $\frac{\rm{hard}-\rm{soft}}{\rm{hard}+\rm{soft}}$, where the soft and hard energy bands are defined as  $E<4$ keV and $E>4$ keV.}}
    \label{fig:hardness}
\end{figure}

Using count rates from the folded profiles, we calculated hardness ratios folded at the \(P = 12.26~\mathrm{min}\) period. The hardness ratio is defined as \(\frac{\mathrm{hard}-\mathrm{soft}}{\mathrm{hard}+\mathrm{soft}}\), where ``soft" and ``hard" are the soft and energy band count rates for the respective telescope. In the XMM-MOS plot (Figure \ref{fig:hardness} left panel), we observe that the hardness ratio anti-correlates with the overall lightcurve data -- for example, the hardness ratio is peak at $\phi\approx 0.5$ where the X-ray flux dips. 

\section{Spectral Analysis} \label{sec:source}

We present phase-averaged and phase-resolved X-ray spectral analysis using \xmm, \nustar\ and \nicer\ data. We jointly fit the \nustar, \xmm\ and \nicer\ in 0.3--30 keV spectra with phenomenological models to constrain the spectral properties. For \nicer\ spectra, we ignored low-energy photons below $1$ keV as the {measured} flux is lower than the flux predicted by \texttt{SCORPEON} background model. {The \nicer\ observations from December 2022 were discarded due to their short GTI's (minutes) and lack of source counts.}
The results of fitting our phenomenological models are summarized in  Table \ref{tab:pheno}. We performed spectral fitting with \texttt{XSPEC} version 12.13.1 \citep{arnaud_xspec_1996}. Each spectral model was multiplied by \texttt{tbabs} to account for the ISM absorption using the Wilms abundance data \citep{wilms_absorption_2000}. In addition, \texttt{constant} was applied as a cross-normalization factor between the different X-ray observations, with respect to the \nicer\ data. 

\subsection{Phase-Averaged Spectral Analysis}

\begin{figure}[htbp]
    \centering
    \subfigure{
    \includegraphics[width=0.47\textwidth]{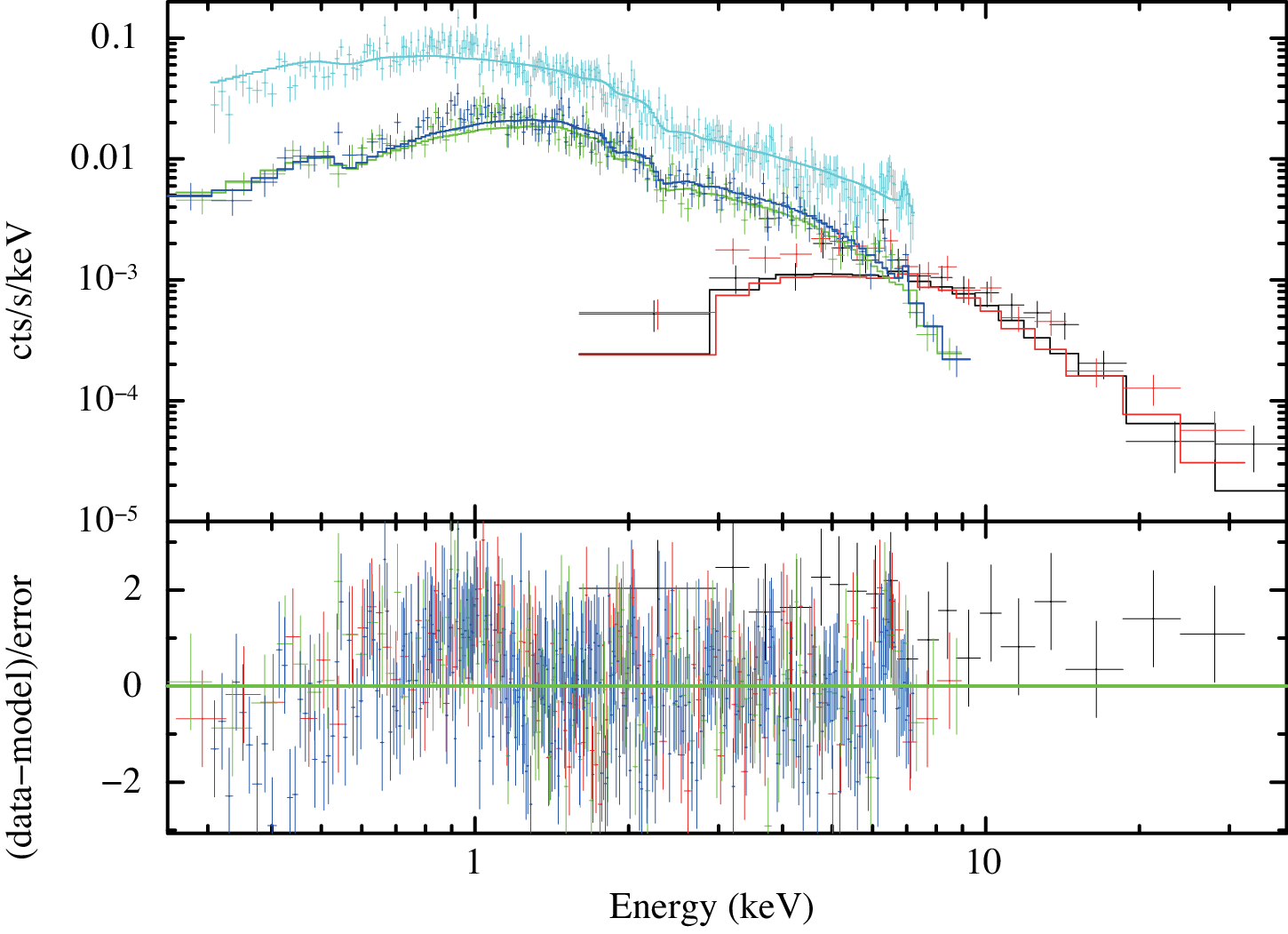}
        \label{fig:first}
    }
    \hspace{0.01\textwidth}
    \subfigure{
        \includegraphics[width=0.47\textwidth]{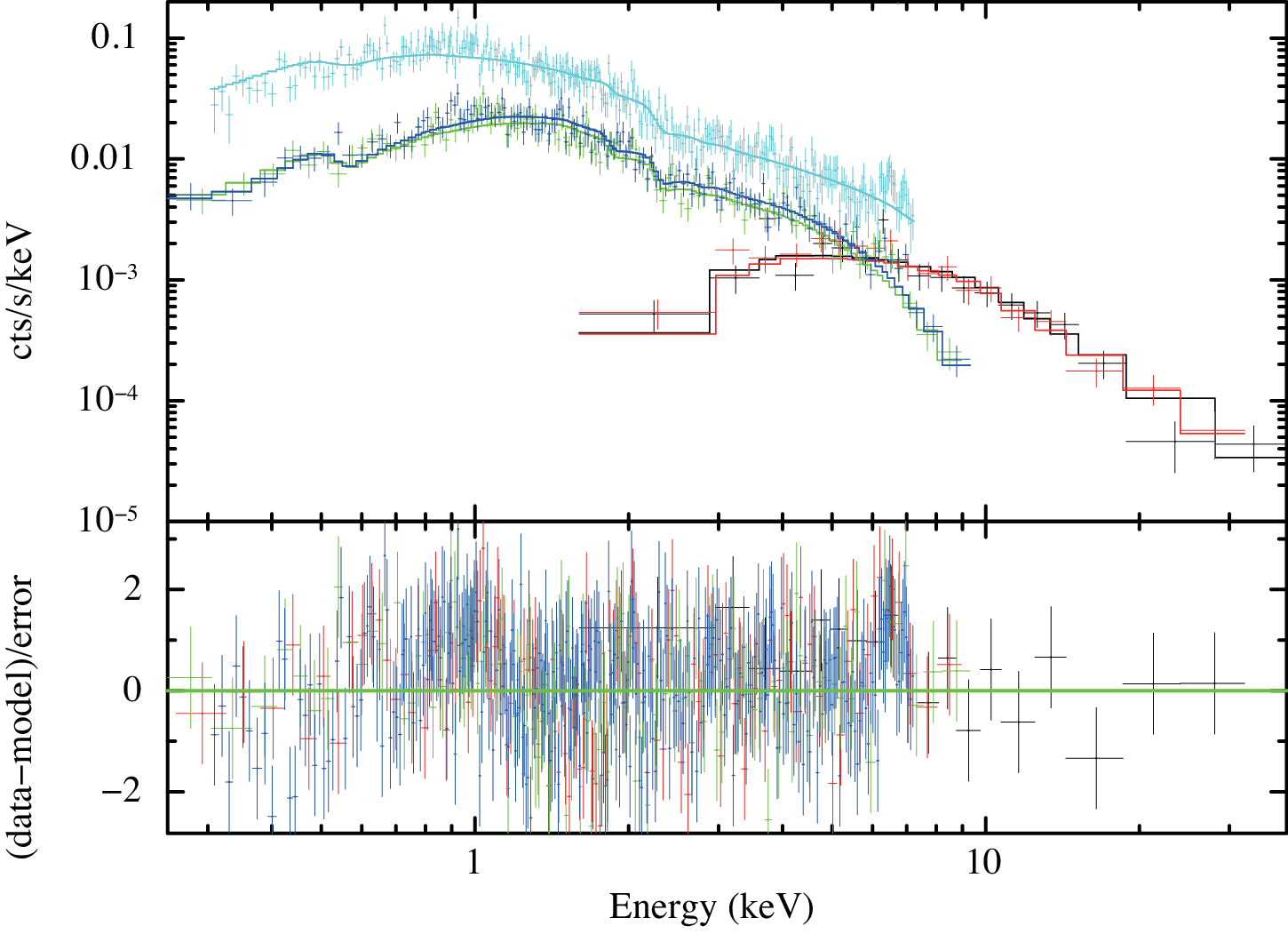}
        \label{fig:second}
    }
    \caption{Joint \textit{XMM-Newton} (light blue: pn; green: MOS1; dark blue: MOS2) and \textit{NuSTAR} (black: FPMA; red: FPMB) spectra fitted by absorbed \texttt{APEC} (left) and absorbed power-law (right) models. Notice the atomic line between 6-7 keV in the residual. \nicer\ spectra are excluded in the figures to highlight the atomic lines but are included in the phenomenological fittings.}
    \label{fig:phenomenologicalspectra}
\end{figure}

First, we fit an absorbed power-law model $\Gamma$ and yielded a fit with $\Gamma=1.4$ and a reduced $\chi^2=1.10$ (Figure \ref{fig:phenomenologicalspectra}).  Large residuals are present near 6--7 keV in \nustar\ and \xmm\ spectra, indicating the presence of atomic lines. To confirm the presence of an atomic line in this regime, we compared the $\chi^2$ of the best-fit models with and without a Gaussian component. Because an analytical F-test is not valid for assessing the significance of a line \citep{2002ApJ...571..545P, 2005ApJ...631.1082M}, we used the Monte Carlo-based \texttt{simftest} command. We simulated $10^4$ spectra drawn under the null hypothesis (assuming that no line is present); only two of the simulated spectra yielded a $\Delta\chi^2$ as large as observed when including a Gaussian model, exhibiting no significant improvement in fit statistic. Thus, the probability of obtaining an improvement in fit as large as observed by chance is $p<2\times10^{-4}$ ($\approx 3.5\sigma$), strongly supporting the presence of a line unaccounted for by a power-law model alone.
We then applied an absorbed optically thin thermal bremsstrahlung model with atomic lines (\texttt{APEC}). The improved spectral fit indicates a plasma temperature $kT\approx 25.6$ keV. A two-temperature model (\texttt{APEC+APEC}) did not improve the fit with $kT_1 \sim 1$ keV and $kT_2 \sim 40$ keV. 
We also added a Gaussian line at 6.4 keV or a blackbody component (\texttt{bbodyrad}) as they are occasionally observed from X-ray spectra of mCVs. The 6.4 keV Gaussian line accounts for the neutral Fe K-$\alpha$ line from X-rays reflected off the WD surface or accretion curtain. This atomic line is most visible in \nustar\ and \xmm\ spectra (Fig \ref{fig:phenomenologicalspectra}). The fit was marginally improved ($\chi^2_\nu=1.07$) with the equivalent Gaussian width (EW = $208.7^{+68.5}_{-71.3}$ eV). The blackbody radiation could arise from the heated polar caps on the WD surface \citep{scaringi_hard_2010}. However, the best-fit blackbody temperature ($kT_{\rm BB} =  3.5$ keV) far exceeded the typical blackbody temperatures for mCVs \citep[$kT_{\rm BB} \simlt0.1$ keV;][]{ramsay_xmm-newton_2002}. In \nicer\ spectra, we found an emission line feature near 0.5 keV that does not have a counterpart in \xmm\ observation. The emission line feature at 0.59 keV is likely caused by solar wind charge exchange (SWCX). Following the guidance from the {\it NICER} helpdesk, we initially added an extra Gaussian line fixed at 0.59 keV to account for the neutral oxygen line when fitting \nicer, \XMM\ and \nustar\ spectra jointly. We later discarded photons below 1 keV altogether, {consistent with our timing analysis,} as \texttt{SCORPEON} model predicts higher background flux than the {measured} flux in that regime. 

\begin{deluxetable}{lccccccc}[ht]
\tablecaption{Phenomenological model fits to X-ray spectra}
\label{tab:xspec_fit}
\tablecolumns{6}
\tablehead{
\colhead{Parameter} &
\colhead{\texttt{pow}} &
\colhead{\texttt{APEC}} &
\colhead{$2\cdot$\texttt{APEC}} &
\colhead{\texttt{APEC+bbody}}&
\colhead{\texttt{APEC+gauss}}&
\colhead{\texttt{APEC+bbody+gauss}}
}
\startdata
$C^a_{XMM}$ & 0.64 & 0.63 & 0.62 & 0.64 & 0.63 &0.62\\
$C^a_{NuSTAR}$ & 0.96 & 1.03 & 0.94 & 0.92 & 1.03 &0.95\\
$N^{(i)}_H [10^{20} \rm{cm}^{-2}]^b\;$ & $8.99\pm 1.0$ & $6.51\pm 0.7$ & $6.77\pm0.9$ & $8.82_{-1.0}^{+0.9}$&$6.66_{-0.8}^{+0.7}$ &$8.62_{-1.8}^{+1.2}$\\
$\Gamma$  & $1.40 \pm 0.04$& .. & .. & .. & .. &..\\
$kT_1$ (keV) & .. & $25.6_{-4.8}^{+7.1}$ & $0.96_{-0.08}^{+0.10}$&  $5.44_{-1.06}^{+2.07}$& $22.8_{-3.45}^{+6.98}$ & $6.16_{-1.72}^{+10.2}$\\
$kT_2$ (keV) & ..   & .. & $42.17_{-10.72}^{+21.83}$ & $3.47_{-0.39}^{+0.58}$ &..&$3.73_{-0.61}^{+3.23}$\\
$Z^c (Z_\odot)$ & ..  & $1.02_{-0.47}^{+0.59}$ & $1.78_{-0.93}^{+1.76}$& $0.16_{-0.10}^{+0.16}$&$0.90_{-0.41}^{+0.57}$&$0.18_{-0.14}^{+0.57}$ \\
$EW^d_{\rm line}$ (eV) & .. & .. & .. & .. & $208.7^{+68.5}_{-71.3}$& $153.0_{-65.8}^{+58.3}$\\
$F_X [10^{-13}$ \fluxcgs]$^e$ & 8.4 & 8.6 & 9.1 & 9.2 &8.8&9.4\\
$\chi^2_{\nu}$ (dof) & 1.10 (763)  & 1.10 (762) & 1.00 (760) & 1.07 (760)&1.07 (761) &1.05 (757)\\
\enddata
\tablecomments{All errors shown are 90\% confidence intervals.}
\vspace{-5pt} 
\tablenotetext{a}{Cross-normalization factor of the \textit{XMM-Newton} and \textit{NuSTAR} data with respect to the \textit{NICER} data.}
\vspace{-5pt} 
\tablenotetext{b}{The ISM hydrogen column density associated with \texttt{tbabs}, which is applied to all models.}
\vspace{-5pt} 
\tablenotetext{c}{Abundance relative to solar.}
\vspace{-5pt} 
\tablenotetext{d}{The equivalent width of the Gaussian component with $E=6.4$ keV and $\sigma=0.01$ keV.}
\vspace{-5pt} 
\tablenotetext{e}{The unabsorbed 3-10 keV flux of the \textit{NICER} data.}
\label{tab:pheno}
\end{deluxetable}

\subsection{Phase-Resolved Spectral Analysis}
\begin{figure}[htbp]
    \centering
    \begin{minipage}{0.47\textwidth}
        \centering
        \includegraphics[width=\textwidth]{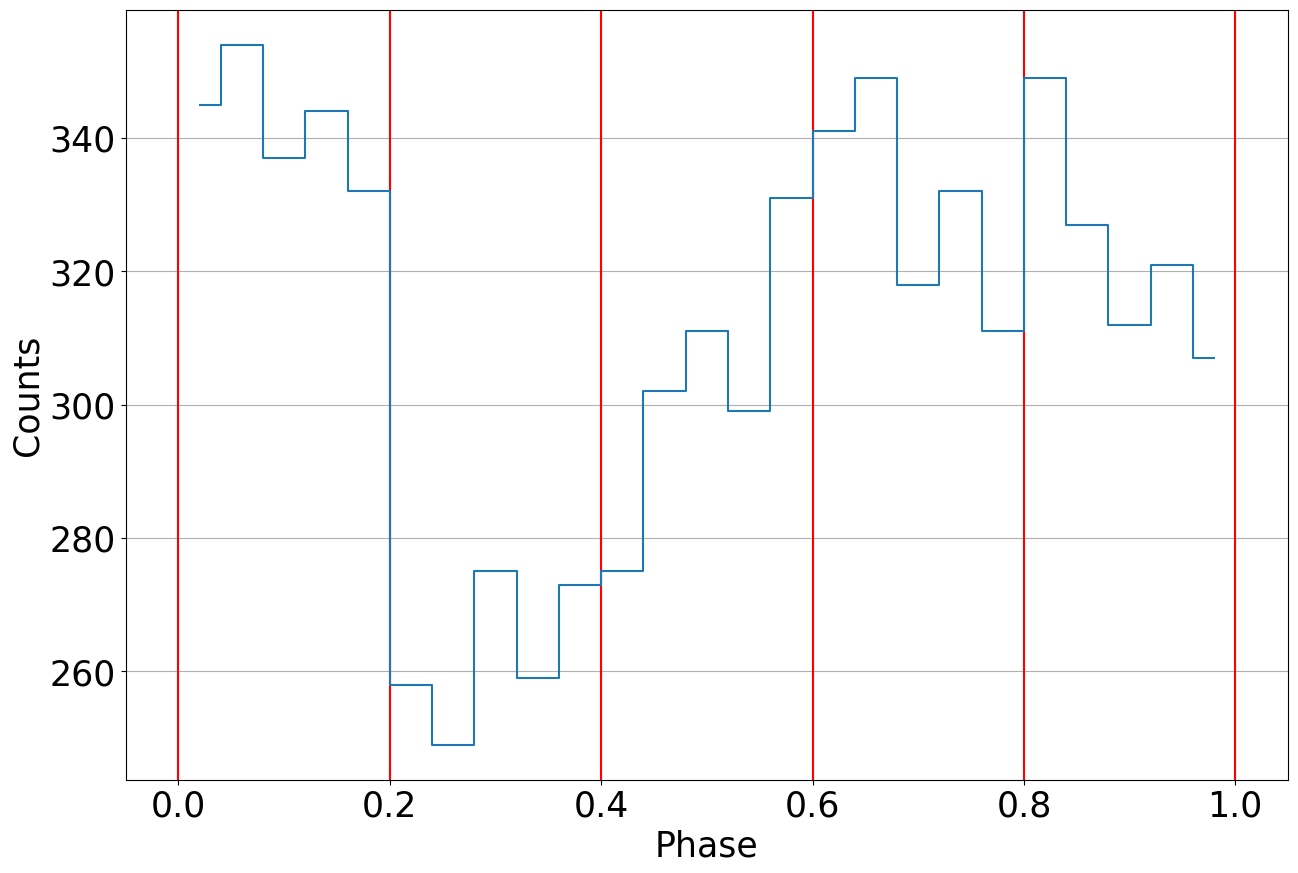}
        \label{fig:image1}
    \end{minipage}
    \hfill
    \begin{minipage}{0.47\textwidth}
        \centering
        \includegraphics[width=\textwidth]{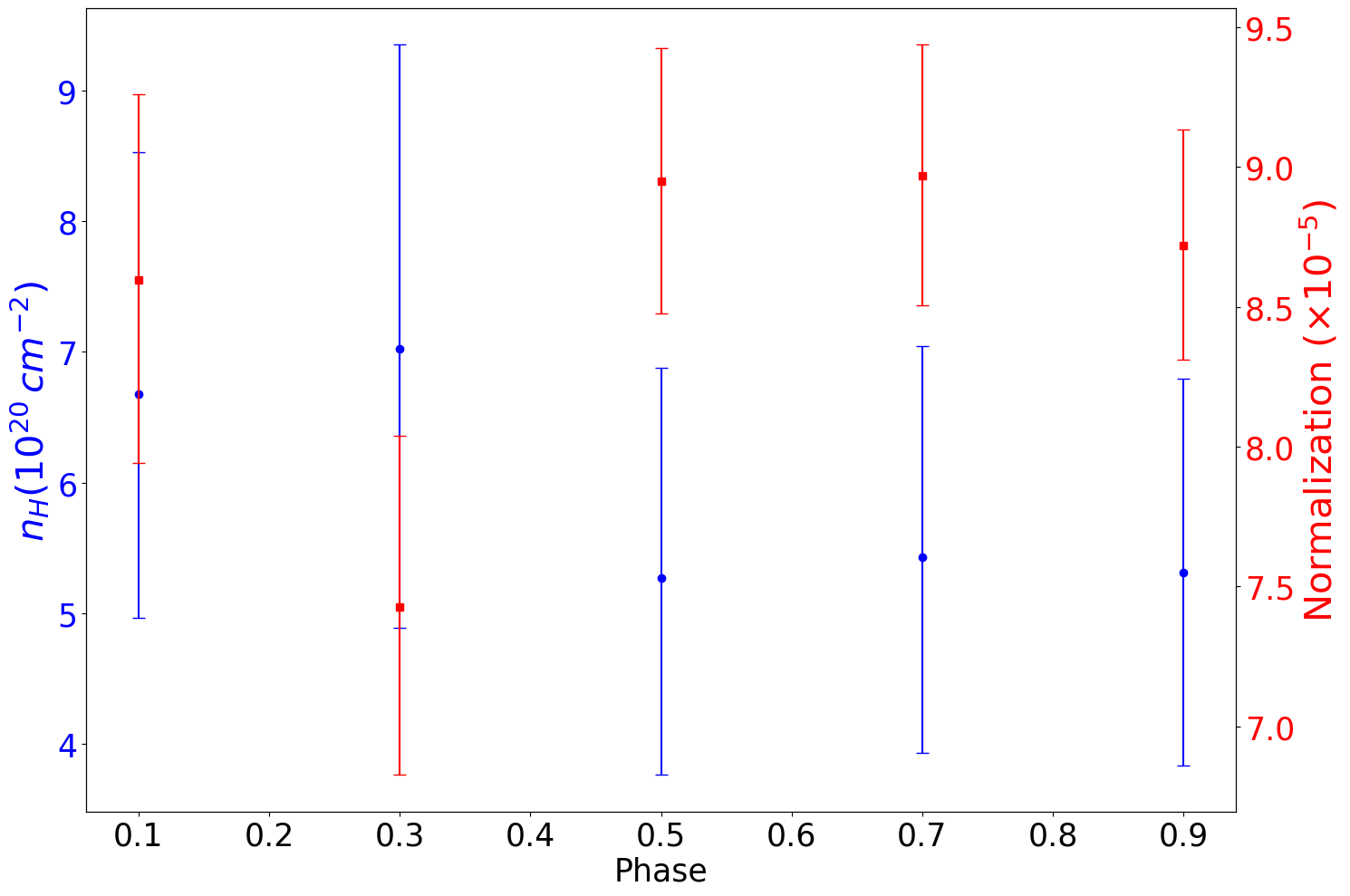}
        \label{fig:image2}
    \end{minipage}
    \caption{Results of phase-resolved spectroscopy. \textit{Left}: The PN camera lightcurve folded at the 12.27 min period, with red lines demarcating the five phase bins ($\phi=0.0-0.2, 0.2-0.4, 0.4-0.6, 0.6-0.8, 0.8-1.0$). \textit{Right}: The variation of the best-fit hydrogen column density $n_H$ and flux normalization across phase bins, with fixed $kT=25.5$ keV and $Z=1.0\,Z_{\odot}$.}
    \label{fig:phase_resolved}
\end{figure}

Given the ample photon statistics, we performed phase-resolved spectral analysis on the \xmm\ data using the X-ray spin period ($P = 12.26$ min) detected in the \xmm\ spectra. 
We extracted \xmm\ spectra from five phase bins with $\Delta\phi=0.2$ intervals and fitted an absorbed \texttt{APEC} model to each phase-resolved spectrum. We fixed the plasma temperature at the phase-averaged best-fit value $kT=25.5$ keV and allowed the ISM hydrogen column density $n_H$ to vary. The fits for all phase bins except the middle bin were excellent with $0.9<\chi^2<1.1,$ while the middle phase bin was subject to poorer fit ($\chi^2>1.2$). We found a small anticorrelation between the X-ray absorption factor quantified by $n_H$ and photon counts (Figure \ref{fig:phase_resolved}), with the highest X-ray absorption corresponding to the bin with the lowest photon counts. Likewise, the flux normalization factor exhibits an anticorrelation with $n_H$ at fixed $kT.$

\section{Discussion}\label{sec:discussion}
We have thoroughly investigated the X-ray properties of {the optical source} ZTF J1851, using \nustar, \xmm\ and \nicer\ data. Below, we discuss the results and implications of our preceding analysis.

\subsection{Source Identification}\label{sec:src_id}

The broadband X-ray spectra of ZTF J1851 display prominent atomic line near 6-7 keV and hard X-ray photons extending to $\sim 30$ keV. The hard X-ray continuum observed by \textit{NuSTAR} rules out {the possibility of being a director accretor like HM Cnc and V407 Vul or the disc-accreting AM CVn's, which have more modest X-ray temperatures} \citep{ramsay_xmm-newton_2005}. 
The presence of prominent atomic lines is inconsistent with featureless power-law X-ray spectra observed in UCXBs \citep{koliopanos_chemical_2021}. 
The X-ray spectra of \src\ are also distinct from much fainter ($L_X \sim 10^{30}$~\lumcgs) and softer ($kT\simlt 8$ keV) X-ray emission observed from the known WD pulsars \citep{Schwope2023}. 

The bright X-ray luminosity ($L_X = 10^{33}$~\lumcgs), hard X-ray spectra, neutral and ionized Fe lines are all typically observed in IPs where the shock-heated accretion curtain emits thermal continuum X-rays and atomic lines. The high plasma temperature of $kT\approx 25$ keV is commonly observed from IPs, whose X-ray spectra are harder than polars or non-magnetic CVs (\cite{mukai_x-ray_2017}, \cite{hailey_evidence_2016}). The neutral Fe fluorescent line, whose equivalent width ($\textup{EW}\sim 200$ eV) is typical among IPs, originates from X-rays reflected off the WD surface or accretion curtain. Knowing ZTF J1851 is an IP, it is also unlikely that the discrepancy between the measured X-ray and optical periods arises from the spin evolution of the WD between the optical period measurement in 2021 and the \xmm\ observation in 2024. The indicated spin derivative of $\dot{P}\sim 10^{-9}s/s,$ is several orders of magnitude higher than typically observed in mCVs \citep{salcedo_broadband_2024}, further validating that the optical period is the one-day alias produced by the Earth rotation. Furthermore, the energy-dependent X-ray modulation supports that the 12.26 min period represents the WD spin. The higher pulsed fractions below 2 keV imply the presence of an accretion curtain that absorbs soft X-rays predominantly. In IPs, as accreting particles are funneled onto WD poles along magnetic field lines from the accretion disk or stream, they form an extended arc-like curtain of infalling gas that leads into the column. As the WD spins, the co-rotating accretion curtain absorbs soft X-rays and obscures the accretion column, causing the energy-dependent spin modulation and phase-dependent spectral hardening (\cite{rosen_exosat_1988}). This is consistent with what we find in the energy-resolved lightcurve and hardness ratio plots {(see Figures \ref{fig:profiles} and \ref{fig:hardness}), where lower-energy lightcurves exhibit stronger modulation with the spin period, a feature commonly seen in IP systems (\cite{joshi_x-ray_2022}; \cite{salcedo_broadband_2024}).}  

\subsection{White Dwarf Mass Determination}\label{sec:mass}

Given that ZTF J1851 is an IP, we proceed to determine its WD mass in this section. 
In mCVs, X-ray emission is typically powered by mass accretion, with the infalling particles converting their gravitational potential energy into electromagnetic radiation via shock-heated gas flow. As the magnetic field disrupts the formation of a complete accretion disk, an accretion flow will be channeled along the magnetic field lines from the innermost accretion disk radius. As the infalling gas reaches supersonic speed, a stand-off shock is formed and heats the gas. Subsequently, the shock-heated gas cools via cyclotron cooling and thermal bremsstrahlung radiation within the accretion column, which emits copious X-rays. Below the stand-off shock, a range of plasma temperatures, densities, and emissivity characterizes X-ray emission from the accretion column. It may be reflected from the fact that the two-temperature \texttt{APEC} model best fits the X-ray spectra as shown in Table \ref{tab:pheno}. 

Following our recent IP papers \citep{salcedo_broadband_2024, vermette_constraining_2023}, we applied our latest 1D accretion spectral model \texttt{MCVSPEC} (Bridges et al. in prep) for modeling X-ray emission from \src. Its application for a polar is presented in \citet{Filor2024}. Our model accounts for the gradient of plasma temperature, density, and X-ray emissivity within the accretion column self-consistently and the effects of X-ray reflection off of the WD surface. The input parameters for \texttt{MCVSPEC} are $M,$ $f$ (fractional accretion area), $L$ (bolometric luminosity [\lumcgs]), $P_{\rm spin}$ (WD spin period [s]), $Z/Z_{\odot}$ (abundance relative to solar), $\cos (i)$ (inclination angle of the reflecting surface), and flux normalization. Our model assumes that the gas particles acquire kinetic energy by falling from a finite magnetospheric radius ($R_m$) to the shock height ($h$). Therefore, the free-fall velocity at the shock height is given by $v_{\rm ff}=\sqrt{2GM(\frac{1}{R+h}-\frac{1}{R_m})}$. This allows us to determine shock temperature ($T_{\rm s}$) and WD mass through the relation $kT_{\rm s}=\frac{3}{8}\mu m_Hv^2_{\rm ff}$ (\cite{aizu_x-ray_1973}; \cite{hayashi_new_2014}; \cite{hailey_evidence_2016}; \cite{suleimanov_gk_2016}; \cite{shaw_measuring_2020}). In addition, some X-rays from the accretion column may be reflected by the WD surface or accretion curtain. These reprocessed X-rays manifest as a neutral Fe fluorescence line at 6.4 keV and a Compton scattering hump above $\sim$ 10 keV. The extent of X-ray reflection depends on the shock height: A tall accretion column diminishes the effects of reflection since the viewing angle of the WD at a higher vantage point would be smaller. We account for this effect by internally implementing the X-ray reflection model (\texttt{reflect}) through the reflection fraction factor ($\Omega/2\pi$), where $\Omega$ is the solid angle of the WD surface viewed from the shock height \citep{magdziarz_angle-dependent_1995}. Our overall spectral model is \texttt{tbabs*(MCVSPEC+gauss)} where the \texttt{reflect} model is implemented in \texttt{MCVSPEC}. 

We estimate the magnetospheric radius using the results of our X-ray timing analysis. As we did not detect a spectral break associated with $R_m$ in the PDS (\cite{suleimanov_gk_2016}), we assume that the innermost accretion disk co-rotates with the WD magnetosphere at $R_m$. Note that most IPs were found to be in spin equilibrium  \citep{patterson_spin-period_2020}. Hence, in the spin equilibrium assumption, we derive $R_m=R_{\rm{co}}=\left(\frac{GMP_{spin}^2}{4\pi^2}\right)^{1/3}$. Since our initial work presented in \citet{vermette_constraining_2023}, we improved our model by internally calculating $R_m$ from spin period to automatically fulfill our spin equilibrium assumption. For each iteration, then, our model uses the estimated mass $M$, $R_m$, and $f$ to compute plasma temperature and density profiles within the accretion column by solving coupled differential equations associated with the mass, momentum and energy conservation. The model keeps varying these parameters ($M, R_m$, and $f$) until the shock height is converged.

\begin{figure}[htbp]
    \centering
    \begin{minipage}{0.7\textwidth}
        \centering
        \includegraphics[width=\textwidth]{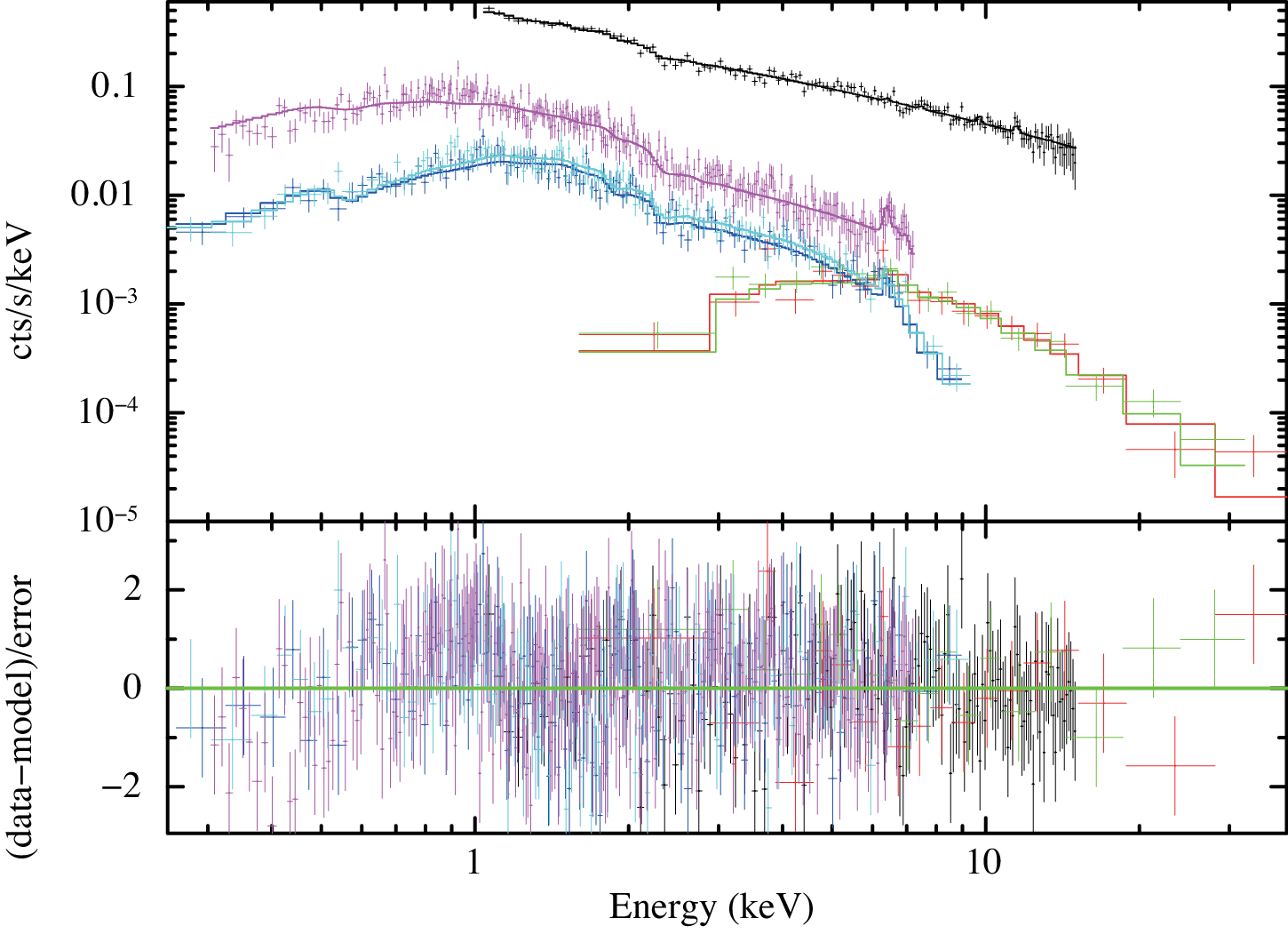}
    \end{minipage}
    \caption{\nicer\ (black), \xmm\ (purple and blue), and \nustar\ (green and red) spectra of \src\ fit by the \texttt{tbabs*(MCVSPEC+gauss)} model with a Gaussian component fixed at $6.4$ keV ($\chi^2_\nu=1.00$).  {For this particular case corresponding to the fractional accretion area fixed at $f=4\times10^{-4},$} the best-fit WD mass is $M=1.25\pm 0.01\,M_{\odot},$ with $\dot{m} = 29.1\,\textup{g}\,\textup{cm}^{-2}\textup{s}^{-1}.$  While the statistical error is quoted for the WD mass measurement above, the systematic error is considered in our final results by considering a range of the fractional accretion area.}
    \label{fig:spec_fit}
\end{figure}

The systematic error associated with our mass measurement model arises from the unknown shock height, which is dependent on the uncertain specific mass accretion rate $\dot{m}=\frac{\dot{M}}{4\pi R^2f},$ where $\dot{M}$ is the total mass accretion rate. At higher specific accretion rates $\dot{m}$, the gas density at the shock height increases, which lowers the speed of sound. As a result, the infalling gas takes longer to decelerate to the shock speed, causing the shock to form closer to the surface---that is, at a lower shock height. We derive the total accretion rate from the bolometric luminosity and WD mass through $L=GM\dot{M}(\frac{1}{R}-\frac{1}{R_m})$. The bolometric luminosity consists of bremmsstralung emission $L_X$ in the X-ray band and cyclotron emission $L_{cyc}$ in the optical, IR, and UV bands. To obtain the X-ray luminosity, we employed the \texttt{cflux} model to find the unabsorbed X-ray flux, which for the \textup{NuSTAR} data is $F_X = (2.6_{-0.3}^{+0.4})\times 10^{-12}$~\fluxcgs. We used a \texttt{constant} factor to calculate the flux for the \textit{XMM-Newton} datasets. Owing to the lack of UV and IR band observations, we estimate $L_{cyc}$ with solely the optical luminosity, using ATLAS o-band and c-band fluxes that sum to $L_{cyc}\approx 2.0\times 10^{-13}$~\fluxcgs\,(Fig \ref{fig:lightcurve}). Using the source distance of $d = 5$ kpc \citep{2021AJ....161..147B}, we determined $L = 8.4\times10^{33}$~\lumcgs\ and $\dot{M}$. 

The unknown fractional accretion column area $f$, however, requires us to consider a range of $f$ values that result in the systematic error associated with the WD mass measurement \citep{salcedo_broadband_2024}. The maximum possible fractional accretion area is $f_{\rm max}=\frac{R}{2R_m}$, assuming a dipole B-field geometry where the accreting gas falls from the entire accretion disk \citep{frank_accretion_2002}. For $M=0.6 M_{\odot}$ and $P_{\rm spin}=12.26$ min, we derived $f_{\rm max}\approx 0.04,$. Higher WD masses would yield $f_{\rm max}\sim0.01\rm{-}0.02$. The lower limit of $f$ value, however, is less constrained. 
In mCVs, soft X-ray blackbody emission often arises from the accretion column base, which is heated by infalling gas particles \citep{scaringi_hard_2010}, and its flux normalization can be used to estimate a $f$ value. However, since we did not detect a blackbody emission component, we set a lower bound of the $f$ value to $10^{-4}$ based on the theoretical estimates proposed by \citet{rosen_exosat_1988}. 

\subsection{X-ray Spectral Fitting Procedure}

The input parameters for \texttt{MCVSPEC} for our optical source ZTF J1851 are generally unknown, with no previously determined magnetic field or inclination angle. Based on our phenomenological model fit results, we fixed the hydrogen column density $n_H$ to $7\times 10^{20} \textup{cm}^{-2}$ for the \texttt{tbabs} model. Following previous X-ray studies (e.g., \cite{hailey_evidence_2016}), the Gaussian component used to model the neutral Fe fluorescence line is centered at $6.4$ keV with fixed $\sigma=0.01$ keV. 

\begin{figure}[htbp]
    \centering
    \begin{minipage}{0.7\textwidth}
        \centering
        \includegraphics[width=\textwidth]{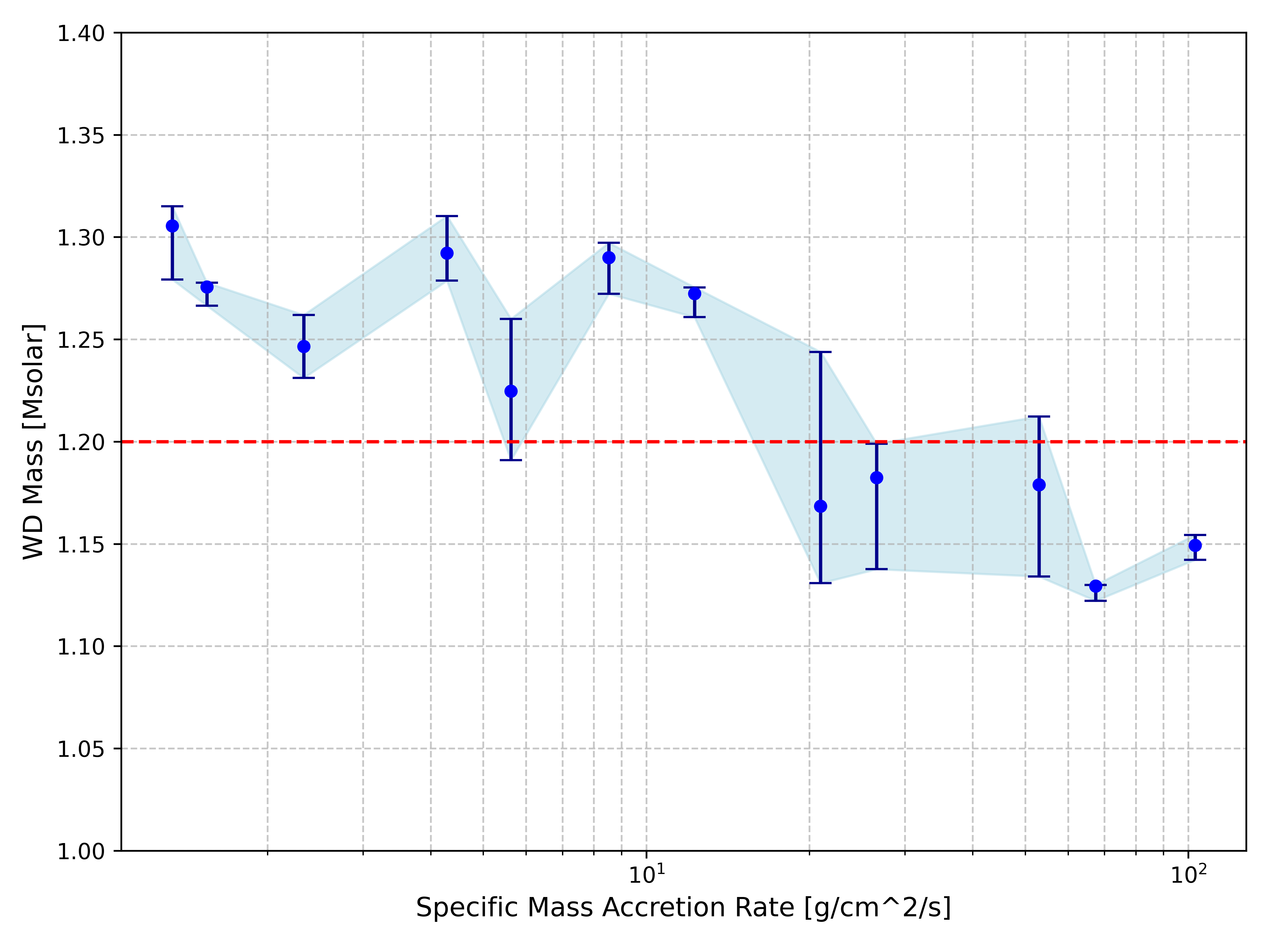}
        \label{fig:image1}
    \end{minipage}
    \caption{The best-fit WD mass as a function of $\dot{m}$ using \texttt{MCVSPEC} with a range of selected $f$ values between $10^{-4}\le f\le 0.01$. In this case, we assumed an initial WD mass of $M_i=1.2M_{\odot}$ (red dashed line) and conducted spectral fitting. All cases yielded a reasonable fit to the X-ray spectra with $\chi_\nu^2 = 1.0\rm{-}1.1$ as shown in Figure \ref{fig:spec_fit}. }
    \label{fig:mcvspec}
\end{figure}

We test a wide range of $\dot{m}$ based on the bolometric luminosity and fractional accretion area. Since the mass accretion rate also depends on the initial mass estimate $M$, we must assume an initial WD mass ($M_i$) for fitting. To ensure our spectral fitting is self-consistent, we compare $M_i$ with the \texttt{MCVSPEC} final mass fit ($M_f$) within their statistical and systematic error range. {We sampled four different initial masses with grid size $\Delta M_i=0.2M_{\odot}$, ranging from $M_i=0.6-1.2M_{\odot}.$} We then iterate through different values of $M_i$ and perform spectral fittings for each case, {and we determined that} lower initial masses at $M<1.2 M_{\odot}$ failed to deliver self-consistent fittings. In Figure \ref{fig:mcvspec}, {which shows the best WD masses for various specific accretion rates for a given initial WD mass value ($M_i=1.2M_{\odot}$ in this case)}, all fittings were performed with an initial guess $M_i=1.2 M_{\odot},$ with $90\%$ error bar associated with each fit plotted. The blue shaded region is the total systematic and statistical error range considering unknown $f$ value. We note that the region intersected by the red line represents the region where the systematic and statistical error of $M_f$ subsumes $M_i$ and is therefore self-consistent. {Considering both the statistical and systematic error,} this yields an estimated WD mass range of $M_{WD}=(1.07-1.32) M_{\odot}.$ Using the magnetic radius formula from \citep{norton_spin_2004}, we further determined {$B\sim 20-40$ MG}. {For descriptions of \texttt{MCVSPEC} fitting procedure applied to other mCVs, see \citet{Filor2024} or \citet{salcedo_broadband_2024}.}

The results of our spectral analysis using \texttt{tbabs*(MCVSPEC+gauss)} are summarized in Table \ref{tab:mcvspec_fit} and Figure \ref{fig:mcvspec} below. Notably, the best-fit WD mass saturates above $M=1.1 M_{\odot}$ for sufficiently large specific accretion rate $\dot{m}$ (Figure \ref{fig:mcvspec}). This saturation phase reflects the fact that as $\dot{m}$ becomes large, the shock height becomes negligible. Consequently, the free-fall velocity $v_{ff}=\sqrt{2GM(\frac{1}{R+h}-\frac{1}{R_m})}$ (as well as plasma temperature and WD mass) becomes insensitive to changes in $h$, leading to a plateau of WD mass estimates. This is a systematic feature of the model rather than a physical phenomenon, and independent estimates of $f$ are needed to place a more stringent constraint on WD mass. It also implies that the WD mass of IP's (or more generally mCV's) with higher specific mass accretion rate can be more tightly constrained. 

\begin{deluxetable*}{lcccccc}[ht]
\color{black}
\tablecaption{Selected {\tt MCVSPEC} fit results to the X-ray spectra}
\label{tab:mcvspec_fit}
\tablecolumns{3}
\tablehead{
\colhead{Parameter}
&
\colhead{Case 1}
&
\colhead{Case 2}
&
\colhead{Case 3}
}
\startdata
$N^{(i)}_H (10^{20} \rm{cm}^{-2})^{a,b}\;$ &7.0& 7.0 & 7.0  \\
$f^b$&$ 10^{-3}$ & $2\times 10^{-3}$&$0.01$\\ 
$\dot{m}_{NICER}$ (g\,cm$^{-2}$\,$s^{-1})^b$ & $10.8$ &  $3.30$&$1.2$ \\
$M (M_{\odot})$& $1.17_{-0.00}^{+0.05}$   & $1.05\pm0.01$&$1.20\pm0.00$ \\
$Z^b (Z_\odot)$ & $0.08_{-0.02}^{+0.01}$&$0.06\pm 0.01 $  & $0.4$ \\
$EW^c_{\rm line}$ (eV) & $163\pm91$  & $168\pm92$ &$166^{+95}_{-90}$  \\
$\cos(i)$ & $0.97_{-0.822}$& $0.97_{-0.822}$ & $0.97_{-0.822}$ \\
$\chi^2_{\nu}$ (dof) &1.00 (756) & 1.06 (760) & 1.11 (760) \\
$R_m/R$ & 39.0&24.7 & 37.0\\
$h/R$ &  3.89\%&7.8\%   & 4.75\% \\
$B$ (MG) &  20.9& 12.5 & 31.9\\
{$kT_{\rm{shock}}$ (keV)}  & 66.6&42.7 & 81.1 &
\enddata
\tablecomments{All errors shown are 90\% confidence intervals. The three cases indicate three distinct fractional accretion area sizes.}
\vspace{-5pt} 
\tablenotetext{a}{The ISM hydrogen column density associated with \texttt{tbabs}, which is applied to all models.}
\vspace{-5pt} 
\tablenotetext{b}{Parameter is frozen.}
\vspace{-5pt} 
\tablenotetext{c}{The equivalent width of the Gaussian component with $E=6.4$ keV and $\sigma=0.01$ keV for \textit{XMM-Newton} data}
\vspace{-5pt} 
\label{tab:mcvspec}
\end{deluxetable*}

Our results using the state-of-the-art spectral model \texttt{MCVSPEC} suggest \src\ possesses a massive and highly magnetized WD with  $M=1.07\rm{-}1.32 M_{\odot}$ and {$B=20\rm{-}40$ MG}. The uncertainties are mainly attributed to the poorly constrained fractional accretion area. This places ZTF J1851 as one of the heavier IPs, above the mean WD mass $\langle M_{WD}\rangle\approx 0.8M_{\odot}$ of known IPs \citep{pala_constraining_2022} and mCVs \citep{shaw_measuring_2020}. For some $f$ values, our fitting suggests that \src\ may contain one of the most massive WDs close to the Chandrasekhar mass limit. The {estimated} magnetic field $B\gtrsim 20$ MG falls above a typical IP's B-field range of 0.1--10 MG \citep{mukai_x-ray_2017}. Note that the B-field range derived by assuming spin equilibrium represents an upper limit because $R_m$ can be smaller than the co-rotation radius $R_{co}$ as pointed out by \citet{2019MNRAS.482.3622S, suleimanov_x-ray_2025}. To consider how $R_m < R_{co}$ affects the WD mass measurement, for a given $f$ value ($f = 10^{-3}$ in this case), we repeated X-ray spectral fitting for a range of $R_m/R_{co} = 0.2\rm{-}1$ and found the WD mass is floored at $M>0.9 M_{\odot}$. The smaller $R_m$ value reduces the free-falling distance between $R_m$ and $h$, thereby lowering the shock temperature. On the other hand, the lower cyclotron cooling rate at lower B-field keeps the overall plasma temperature higher in the accretion column. These counteracting effects lead to the minimum WD mass of $M = 0.9M_{\odot}$, being reduced from the lower bound at $M=1.0M_\odot$ (when $R_m = R_{co}$), at $R_m/R_{co} \approx 0.6$, corresponding to $B\sim 1$ MG. Therefore, if $R_m < R_{co}$ is assumed, the WD mass estimate could be lower by $\sim10$\% and this systematic effect will be addressed in our forthcoming {\tt MCVSPEC} model paper (Bridges et al. in prep). 

\section{Conclusion}\label{sec:conclusion}

We report the detection of a hard X-ray counterpart for \src\ and present the first investigation of the X-ray temporal and spectral properties of ZTF J1851. The broadband X-ray observations led to determining the intrinsic spin period and identifying the source as a massive IP with $M_{\rm WD} > 1 M_\odot${, assuming spin equilibrium}. A combination of the soft and hard X-ray telescope data obtained by \xmm, \nicer\ and \nustar\ enabled the measurements of the spin period and WD mass. {Our spectral model \texttt{MCVSPEC} is the most self-consistent model that accounts for physical effects such as finite magnetospheric radius and X-ray reflection from the WD surface. Future observations of the source could improve the systematic uncertainty of our measurements by constraining $f$ if they can detect a blackbody component, as well as better measuring other WD properties like mass and spin period derivative.}

We also report six short (approximately day long) optical outbursts
over a course of nearly 7 years. The outbursts are much shorter in
duration than typical dwarf nova outbursts but similar in duration to
those due to magnetic gating or from micronovae events (see
\citet{2024ApJ...962L..34I} for a discussion). 
If we could confirm the day long duration of optical outbursts in \src using multiple all-sky surveys or TESS data, then this would
provide evidence for these bursts being caused by magnetic gating and
therefore for the WD having a significant magnetic field. With a growing number of periodic sources discovered by optical and X-ray all-sky surveys (e.g., \citet{Mondal2024, Schwope2024}), our results highlight the importance of follow-up broadband X-ray observations to identify periodic optical sources in the future. Similar X-ray follow-up observations could identify new CVs and UCBs through the current and future optical surveys \citep{Rodriguez2024}. 

\section{Acknowledgments}
Support for this work was provided by NASA through \nicer\ Cycle 4 (NNH21ZDA001N-NICER), \xmm\ Cycle 22 (XMMNC22) Guest Observer program, and NASA ADAP program (NNH22ZDA001N-ADAP) grants. We thank Ciro Salcedo for his helpful discussions and contributions on timing analysis. GOTO (https://goto-observatory.org) is a network of telescopes that is
principally funded by the STFC and operated at the Roque de los Muchachos Observatory on La Palma, Spain, and Siding Spring
Observatory in NSW, Australia, on behalf of a consortium including the
University of Warwick, Monash University, Armagh Observatory $\&$ Planetarium, the University of Leicester, the University of Sheffield,
the National Astronomical Research Institute of Thailand (NARIT), the
University of Turku, the University of Portsmouth, the University of
Manchester and the Instituto de Astrofisica de Canarias (IAC).

\bibliographystyle{aasjournal}
\bibliography{bibliography}

\begin{thebibliography}{}
\expandafter\ifx\csname natexlab\endcsname\relax\def\natexlab#1{#1}\fi
\providecommand{\url}[1]{\href{#1}{#1}}
\providecommand{\dodoi}[1]{doi:~\href{http://doi.org/#1}{\nolinkurl{#1}}}
\providecommand{\doeprint}[1]{\href{http://ascl.net/#1}{\nolinkurl{http://ascl.net/#1}}}
\providecommand{\doarXiv}[1]{\href{https://arxiv.org/abs/#1}{\nolinkurl{https://arxiv.org/abs/#1}}}

\bibitem[{Aizu(1973)}]{aizu_x-ray_1973}
Aizu, K. 1973, 49, 1184, \dodoi{10.1143/PTP.49.1184}

\bibitem[{Arnaud(1996)}]{arnaud_xspec_1996}
Arnaud, K.~A. 1996, 101, 17.
\newblock \url{https://ui.adsabs.harvard.edu/abs/1996ASPC..101...17A}

\bibitem[{Bachetti(2018)}]{bachetti_hendrics_2018}
Bachetti, M. 2018, ascl:1805.019.
\newblock \url{https://ui.adsabs.harvard.edu/abs/2018ascl.soft05019B}

\bibitem[{Bachetti {et~al.}(2024)Bachetti, Huppenkothen, Stevens, Swinbank, Mastroserio, Lucchini, Lai, Buchner, Desai, Joshi, Pisanu, Pisupati, Sharma, Tripathi, \& Vats}]{bachetti_stingray_2024}
Bachetti, M., Huppenkothen, D., Stevens, A., {et~al.} 2024, 9, 7389, \dodoi{10.21105/joss.07389}

\bibitem[{{Bailer-Jones} {et~al.}(2021){Bailer-Jones}, {Rybizki}, {Fouesneau}, {Demleitner}, \& {Andrae}}]{2021AJ....161..147B}
{Bailer-Jones}, C.~A.~L., {Rybizki}, J., {Fouesneau}, M., {Demleitner}, M., \& {Andrae}, R. 2021, \aj, 161, 147, \dodoi{10.3847/1538-3881/abd806}

\bibitem[{Bellm {et~al.}(2019)Bellm, Kulkarni, Graham, Dekany, Smith, Riddle, Masci, Helou, Prince, Adams, Barbarino, Barlow, Bauer, Beck, Belicki, Biswas, Blagorodnova, Bodewits, Bolin, Brinnel, Brooke, Bue, Bulla, Burruss, Cenko, Chang, Connolly, Coughlin, Cromer, Cunningham, De, Delacroix, Desai, Duev, Eadie, Farnham, Feeney, Feindt, Flynn, Franckowiak, Frederick, Fremling, Gal-Yam, Gezari, Giomi, Goldstein, Golkhou, Goobar, Groom, Hacopians, Hale, Henning, Ho, Hover, Howell, Hung, Huppenkothen, Imel, Ip, Ivezić, Jackson, Jones, Juric, Kasliwal, Kaspi, Kaye, Kelley, Kowalski, Kramer, Kupfer, Landry, Laher, Lee, Lin, Lin, Lunnan, Giomi, Mahabal, Mao, Miller, Monkewitz, Murphy, Ngeow, Nordin, Nugent, Ofek, Patterson, Penprase, Porter, Rauch, Rebbapragada, Reiley, Rigault, Rodriguez, van Roestel, Rusholme, van Santen, Schulze, Shupe, Singer, Soumagnac, Stein, Surace, Sollerman, Szkody, Taddia, Terek, Van~Sistine, van Velzen, Vestrand, Walters, Ward, Ye, Yu, Yan, \& Zolkower}]{bellm_zwicky_2019}
Bellm, E.~C., Kulkarni, S.~R., Graham, M.~J., {et~al.} 2019, 131, 018002, \dodoi{10.1088/1538-3873/aaecbe}

\bibitem[{Burdge {et~al.}(2022)Burdge, Marsh, Fuller, Bellm, Caiazzo, Chakrabarty, Coughlin, De, Dhillon, Graham, Rodríguez-Gil, Jaodand, Kaplan, Kara, Kong, Kulkarni, Li, Littlefair, Majid, Mróz, Pearlman, Phinney, Roestel, Simcoe, Andreoni, Drake, Dekany, Duev, Kool, Mahabal, Medford, Riddle, \& Prince}]{burdge_62-minute_2022}
Burdge, K.~B., Marsh, T.~R., Fuller, J., {et~al.} 2022, 605, 41, \dodoi{10.1038/s41586-022-04551-1}

\bibitem[{{Dyer} {et~al.}(2024){Dyer}, {Ackley}, {Jim{\'e}nez-Ibarra}, {Lyman}, {Ulaczyk}, {Steeghs}, {Galloway}, {Dhillon}, {O'Brien}, {Ramsay}, {Noysena}, {Kotak}, {Breton}, {Nuttall}, {Pall{\'e}}, {Pollacco}, {Killestein}, {Kumar}, {O'Neill}, {Kelsey}, {Godson}, \& {Jarvis}}]{2024SPIE13094E..1XD}
{Dyer}, M.~J., {Ackley}, K., {Jim{\'e}nez-Ibarra}, F., {et~al.} 2024, in Society of Photo-Optical Instrumentation Engineers (SPIE) Conference Series, Vol. 13094, Ground-based and Airborne Telescopes X, ed. H.~K. {Marshall}, J.~{Spyromilio}, \& T.~{Usuda}, 130941X, \dodoi{10.1117/12.3018305}

\bibitem[{{Filor} {et~al.}(2024){Filor}, {Mori}, {Bridges}, {Hailey}, {Buckley}, {Ramsay}, {Schwope}, {Suleimanov}, {Wolff}, \& {Wood}}]{Filor2024}
{Filor}, L.~W., {Mori}, K., {Bridges}, G., {et~al.} 2024, arXiv e-prints, arXiv:2412.11273, \dodoi{10.48550/arXiv.2412.11273}

\bibitem[{Frank {et~al.}(2002)Frank, King, \& Raine}]{frank_accretion_2002}
Frank, J., King, A., \& Raine, D.~J. 2002, Accretion Power in Astrophysics: Third Edition.
\newblock \url{https://ui.adsabs.harvard.edu/abs/2002apa..book.....F}

\bibitem[{Hailey {et~al.}(2016)Hailey, Mori, Perez, Canipe, Hong, Tomsick, Boggs, Christensen, Craig, Fornasini, Grindlay, Harrison, Nynka, Rahoui, Stern, Zhang, \& Zhang}]{hailey_evidence_2016}
Hailey, C.~J., Mori, K., Perez, K., {et~al.} 2016, 826, 160, \dodoi{10.3847/0004-637X/826/2/160}

\bibitem[{Harrison {et~al.}(2013)Harrison, Craig, Christensen, Hailey, Zhang, Boggs, Stern, Cook, Forster, Giommi, Grefenstette, Kim, Kitaguchi, Koglin, Madsen, Mao, Miyasaka, Mori, Perri, Pivovaroff, Puccetti, Rana, Westergaard, Willis, Zoglauer, An, Bachetti, Barrière, Bellm, Bhalerao, Brejnholt, Fuerst, Liebe, Markwardt, Nynka, Vogel, Walton, Wik, Alexander, Cominsky, Hornschemeier, Hornstrup, Kaspi, Madejski, Matt, Molendi, Smith, Tomsick, Ajello, Ballantyne, Baloković, Barret, Bauer, Blandford, Brandt, Brenneman, Chiang, Chakrabarty, Chenevez, Comastri, Dufour, Elvis, Fabian, Farrah, Fryer, Gotthelf, Grindlay, Helfand, Krivonos, Meier, Miller, Natalucci, Ogle, Ofek, Ptak, Reynolds, Rigby, Tagliaferri, Thorsett, Treister, \& Urry}]{harrison_nuclear_2013}
Harrison, F.~A., Craig, W.~W., Christensen, F.~E., {et~al.} 2013, 770, 103, \dodoi{10.1088/0004-637X/770/2/103}

\bibitem[{Hayashi \& Ishida(2014)}]{hayashi_new_2014}
Hayashi, T., \& Ishida, M. 2014, 438, 2267, \dodoi{10.1093/mnras/stt2342}

\bibitem[{Huppenkothen {et~al.}(2019)Huppenkothen, Bachetti, Stevens, Migliari, Balm, Hammad, Khan, Mishra, Rashid, Sharma, Martinez~Ribeiro, \& Valles~Blanco}]{huppenkothen_stingray_2019}
Huppenkothen, D., Bachetti, M., Stevens, A.~L., {et~al.} 2019, 881, 39, \dodoi{10.3847/1538-4357/ab258d}

\bibitem[{{I{\l}kiewicz} {et~al.}(2024){I{\l}kiewicz}, {Scaringi}, {Veresvarska}, {De Martino}, {Littlefield}, {Knigge}, {Paice}, \& {Sahu}}]{2024ApJ...962L..34I}
{I{\l}kiewicz}, K., {Scaringi}, S., {Veresvarska}, M., {et~al.} 2024, \apjl, 962, L34, \dodoi{10.3847/2041-8213/ad243c}

\bibitem[{Joshi {et~al.}(2022)Joshi, Wang, Pandey, Singh, Naik, Raj, Anupama, \& Rawat}]{joshi_x-ray_2022}
Joshi, A., Wang, W., Pandey, J.~C., {et~al.} 2022, 657, A12, \dodoi{10.1051/0004-6361/202142193}

\bibitem[{Kato \& Kojiguchi(2021)}]{kato_ztf_2021}
Kato, T., \& Kojiguchi, N. 2021, {ZTF} J185139.81+171430.3 = {ZTF}18abnbzvx: the second white dwarf pulsar?,  {arXiv}

\bibitem[{Koliopanos {et~al.}(2021)Koliopanos, Péault, Vasilopoulos, \& Webb}]{koliopanos_chemical_2021}
Koliopanos, F., Péault, M., Vasilopoulos, G., \& Webb, N. 2021, 501, 548, \dodoi{10.1093/mnras/staa3474}

\bibitem[{Lomb(1976)}]{lomb_least-squares_1976}
Lomb, N.~R. 1976, 39, 447, \dodoi{10.1007/BF00648343}

\bibitem[{Magdziarz \& Zdziarski(1995)}]{magdziarz_angle-dependent_1995}
Magdziarz, P., \& Zdziarski, A.~A. 1995, 273, 837, \dodoi{10.1093/mnras/273.3.837}

\bibitem[{{Mondal} {et~al.}(2024){Mondal}, {Ponti}, {Bao}, {Haberl}, {Campana}, {Hailey}, {Mandel}, {Mereghetti}, {Mori}, {Morris}, {Rea}, \& {Sidoli}}]{Mondal2024}
{Mondal}, S., {Ponti}, G., {Bao}, T., {et~al.} 2024, \aap, 686, A125, \dodoi{10.1051/0004-6361/202449527}

\bibitem[{{Mori} {et~al.}(2005){Mori}, {Chonko}, \& {Hailey}}]{2005ApJ...631.1082M}
{Mori}, K., {Chonko}, J.~C., \& {Hailey}, C.~J. 2005, \apj, 631, 1082, \dodoi{10.1086/432632}

\bibitem[{Mukai(2017)}]{mukai_x-ray_2017}
Mukai, K. 2017, 129, 062001, \dodoi{10.1088/1538-3873/aa6736}

\bibitem[{Norton {et~al.}(2004)Norton, Wynn, \& Somerscales}]{norton_spin_2004}
Norton, A.~J., Wynn, G.~A., \& Somerscales, R.~V. 2004, 614, 349, \dodoi{10.1086/423333}

\bibitem[{Pala {et~al.}(2022)Pala, Gänsicke, Belloni, Parsons, Marsh, Schreiber, Breedt, Knigge, Sion, Szkody, Townsley, Bildsten, Boyd, Cook, De Martino, Godon, Kafka, Kouprianov, Long, Monard, Myers, Nelson, Nogami, Oksanen, Pickard, Poyner, Reichart, Rodriguez Perez, Shears, Stubbings, \& Toloza}]{pala_constraining_2022}
Pala, A.~F., Gänsicke, B.~T., Belloni, D., {et~al.} 2022, 510, 6110, \dodoi{10.1093/mnras/stab3449}

\bibitem[{Patterson {et~al.}(2020)Patterson, Miguel, Kemp, Dvorak, Monard, Hambsch, Vanmunster, Skillman, Cejudo, Campbell, Roberts, Jones, Cook, Bolt, Rea, Ulowetz, Krajci, Menzies, Lowther, Goff, Stein, Wood, Myers, Stone, Uthas, Karamehmetoglu, Seargeant, \& {McCormick}}]{patterson_spin-period_2020}
Patterson, J., Miguel, E.~d., Kemp, J., {et~al.} 2020, 897, 70, \dodoi{10.3847/1538-4357/ab863d}

\bibitem[{Pietrukowicz {et~al.}(2019)Pietrukowicz, Mróz, Udalski, Soszyński, \& Skowron}]{pietrukowicz_discovery_2019}
Pietrukowicz, P., Mróz, P., Udalski, A., Soszyński, I., \& Skowron, J. 2019, 881, L41, \dodoi{10.3847/2041-8213/ab372d}

\bibitem[{{Protassov} {et~al.}(2002){Protassov}, {van Dyk}, {Connors}, {Kashyap}, \& {Siemiginowska}}]{2002ApJ...571..545P}
{Protassov}, R., {van Dyk}, D.~A., {Connors}, A., {Kashyap}, V.~L., \& {Siemiginowska}, A. 2002, \apj, 571, 545, \dodoi{10.1086/339856}

\bibitem[{Ramsay \& Cropper(2002)}]{ramsay_xmm-newton_2002}
Ramsay, G., \& Cropper, M. 2002, 334, 805, \dodoi{10.1046/j.1365-8711.2002.05536.x}

\bibitem[{Ramsay {et~al.}(2005)Ramsay, Hakala, Wu, Cropper, Mason, Córdova, \& Priedhorsky}]{ramsay_xmm-newton_2005}
Ramsay, G., Hakala, P., Wu, K., {et~al.} 2005, 357, 49, \dodoi{10.1111/j.1365-2966.2005.08574.x}

\bibitem[{{Rodriguez} {et~al.}(2024){Rodriguez}, {El-Badry}, {Suleimanov}, {Pala}, {Kulkarni}, {Gaensicke}, {Mori}, {Rich}, {Sarkar}, {Bao}, {Lopes de Oliveira}, {Ramsay}, {Szkody}, {Graham}, {Prince}, {Caiazzo}, {Vanderbosch}, {van Roestel}, {Das}, {Qin}, {Kasliwal}, {Wold}, {Groom}, {Reiley}, \& {Riddle}}]{Rodriguez2024}
{Rodriguez}, A.~C., {El-Badry}, K., {Suleimanov}, V., {et~al.} 2024, arXiv e-prints, arXiv:2408.16053, \dodoi{10.48550/arXiv.2408.16053}

\bibitem[{Rosen {et~al.}(1988)Rosen, Mason, \& Córdova}]{rosen_exosat_1988}
Rosen, S.~R., Mason, K.~O., \& Córdova, F.~A. 1988, 231, 549, \dodoi{10.1093/mnras/231.3.549}

\bibitem[{Salcedo {et~al.}(2024)Salcedo, Mori, Bridges, Hailey, Buckley, de~Oliveira, Ramsay, \& van Dyk}]{salcedo_broadband_2024}
Salcedo, C., Mori, K., Bridges, G., {et~al.} 2024, A Broadband X-ray Investigation of Fast-Spinning Intermediate Polar {CTCV} J2056-3014,  {arXiv}, \dodoi{10.48550/arXiv.2409.18247}

\bibitem[{Scargle(1982)}]{scargle_studies_1982}
Scargle, J.~D. 1982, 263, 835, \dodoi{10.1086/160554}

\bibitem[{Scaringi {et~al.}(2010)Scaringi, Bird, Norton, Knigge, Hill, Clark, Dean, {McBride}, Barlow, Bassani, Bazzano, Fiocchi, \& Landi}]{scaringi_hard_2010}
Scaringi, S., Bird, A.~J., Norton, A.~J., {et~al.} 2010, 401, 2207, \dodoi{10.1111/j.1365-2966.2009.15826.x}

\bibitem[{{Schwope} {et~al.}(2023){Schwope}, {Marsh}, {Standke}, {Pelisoli}, {Potter}, {Buckley}, {Munday}, \& {Dhillon}}]{Schwope2023}
{Schwope}, A., {Marsh}, T.~R., {Standke}, A., {et~al.} 2023, \aap, 674, L9, \dodoi{10.1051/0004-6361/202346589}

\bibitem[{{Schwope} {et~al.}(2024){Schwope}, {Knauff}, {Kurpas}, {Salvato}, {Stelzer}, {St{\"u}tz}, \& {Tub{\'\i}n-Arenas}}]{Schwope2024}
{Schwope}, A.~D., {Knauff}, K., {Kurpas}, J., {et~al.} 2024, \aap, 690, A243, \dodoi{10.1051/0004-6361/202450537}

\bibitem[{Shaw {et~al.}(2020)Shaw, Heinke, Mukai, Tomsick, Doroshenko, Suleimanov, Buisson, Gandhi, Grefenstette, Hare, Jiang, Ludlam, Rana, \& Sivakoff}]{shaw_measuring_2020}
Shaw, A.~W., Heinke, C.~O., Mukai, K., {et~al.} 2020, 498, 3457, \dodoi{10.1093/mnras/staa2592}

\bibitem[{Steeghs {et~al.}(2022)Steeghs, Galloway, Ackley, Dyer, Lyman, Ulaczyk, Cutter, Mong, Dhillon, O’Brien, Ramsay, Poshyachinda, Kotak, Nuttall, Pallé, Breton, Pollacco, Thrane, Aukkaravittayapun, Awiphan, Burhanudin, Chote, Chrimes, Daw, Duffy, Eyles-Ferris, Gompertz, Heikkilä, Irawati, Kennedy, Killestein, Kuncarayakti, Levan, Littlefair, Makrygianni, Marsh, Mata-Sanchez, Mattila, Maund, {McCormac}, Mkrtichian, Mullaney, Noysena, Patel, Rol, Sawangwit, Stanway, Starling, Strøm, Tooke, West, White, \& Wiersema}]{steeghs_gravitational-wave_2022}
Steeghs, D., Galloway, D.~K., Ackley, K., {et~al.} 2022, 511, 2405, \dodoi{10.1093/mnras/stac013}

\bibitem[{Suleimanov {et~al.}(2016)Suleimanov, Doroshenko, Ducci, Zhukov, \& Werner}]{suleimanov_gk_2016}
Suleimanov, V., Doroshenko, V., Ducci, L., Zhukov, G.~V., \& Werner, K. 2016, 591, A35, \dodoi{10.1051/0004-6361/201628301}

\bibitem[{{Suleimanov} {et~al.}(2019){Suleimanov}, {Doroshenko}, \& {Werner}}]{2019MNRAS.482.3622S}
{Suleimanov}, V.~F., {Doroshenko}, V., \& {Werner}, K. 2019, \mnras, 482, 3622, \dodoi{10.1093/mnras/sty2952}

\bibitem[{Suleimanov {et~al.}(2025)Suleimanov, Ducci, Doroshenko, \& Werner}]{suleimanov_x-ray_2025}
Suleimanov, V.~F., Ducci, L., Doroshenko, V., \& Werner, K. 2025, 700, A180, \dodoi{10.1051/0004-6361/202554368}

\bibitem[{Takata {et~al.}(2021)Takata, Wang, Wang, Lin, Hu, Li, \& Kong}]{takata_x-ray_2021}
Takata, J., Wang, X.~F., Wang, H.~H., {et~al.} 2021, 907, 115, \dodoi{10.3847/1538-4357/abd0f8}

\bibitem[{Tonry {et~al.}(2018)Tonry, Denneau, Flewelling, Heinze, Onken, Smartt, Stalder, Weiland, \& Wolf}]{tonry_atlas_2018}
Tonry, J.~L., Denneau, L., Flewelling, H., {et~al.} 2018, 867, 105, \dodoi{10.3847/1538-4357/aae386}

\bibitem[{Vermette {et~al.}(2023)Vermette, Salcedo, Mori, Gerber, Yoon, Bridges, Hailey, Haberl, Hong, Grindlay, Ponti, \& Ramsay}]{vermette_constraining_2023}
Vermette, B., Salcedo, C., Mori, K., {et~al.} 2023, 954, 138, \dodoi{10.3847/1538-4357/ace90c}

\bibitem[{Wilms {et~al.}(2000)Wilms, Allen, \& {McCray}}]{wilms_absorption_2000}
Wilms, J., Allen, A., \& {McCray}, R. 2000, 542, 914, \dodoi{10.1086/317016}

\end{thebibliography}

\end{document}